\documentclass[letterpaper,11pt]{article}

\usepackage{tikz}
\usetikzlibrary{shapes,calc,arrows,patterns,decorations.pathmorphing
	,decorations.markings}
      \usetikzlibrary{arrows.meta}

\usepackage[english]{babel}
\usepackage{theorem}
\theoremheaderfont{\itshape\bfseries}
{\theorembodyfont{\itshape}
	\newtheorem{assumption}{\textbf{Assumption}}
	\newtheorem{lemma}{\textbf{Lemma}}
	\newtheorem{definition}{\textbf{Definition}}
	\newtheorem{theorem}{\textbf{Theorem}}
	
	\newtheorem{remark}{\textbf{Remark}}
	
	\newtheorem{problem}{\textbf{Problem}}
}

\usepackage{cite}
\usepackage{amssymb}
\usepackage{algorithmic}
\usepackage{newtxtext,newtxmath}
\usepackage{amsmath}
\usepackage{amsfonts}
\usepackage{mathrsfs}
\usepackage[ruled,vlined]{algorithm2e}
\usepackage[most]{tcolorbox}

\newcommand{\T}{^{\mbox{\tiny T}}}
\newcommand{\R}{\mathbb{R}}
\newcommand{\C}{\mathbb{C}}

\let\geq\geqslant

\newenvironment{proof}[1][Proof]%
{\par\noindent\textit{#1:\ }}%
{\hspace*{\fill} \rule{6pt}{6pt}}
\newenvironment{proof*}[1][Proof]%
{\par\noindent\textit{#1:\ }}{}

\DeclareMathOperator{\diag}{diag}

\DeclareMathOperator{\rank}{rank}

\newenvironment{system}[1]%
{\setlength{\arraycolsep}{0.5mm}\left\{ \; \begin{array}{#1}}%
	{\end{array} \right.}
\newenvironment{system*}[1]%
{\setlength{\arraycolsep}{0.5mm} \begin{array}{#1}}%
	{\end{array}}

\begin{document}

\title{Scalable Exact Output Synchronization of Discrete-Time Multi-Agent Systems in the Presence of Disturbances and Measurement Noise With Known Frequencies} 
\author{Zhenwei Liu, Meirong Zhang, Ali Saberi,
  and Anton A. Stoorvogel
  \thanks{Zhenwei Liu is with College of Information Science and
    Engineering, Northeastern University, Shenyang 110819,
    China (e-mail: liuzhenwei@ise.neu.edu.cn)} 
      \thanks{Meirong Zhang is with
    	School of Engineering and Applied Science, Gonzaga University, Spokane, WA 99258, USA (e-mail: zhangm@gonzaga.edu)}
  \thanks{Ali Saberi is with
    School of Electrical Engineering and Computer Science, Washington
    State University, Pullman, WA 99164, USA (e-mail: saberi@wsu.edu)}
  \thanks{Anton A. Stoorvogel is with Department of Electrical
    Engineering, Mathematics and Computer Science, University of
    Twente, P.O. Box 217, Enschede, The Netherlands (e-mail:
    A.A.Stoorvogel@utwente.nl)}} 

\maketitle

\begin{abstract}
This paper aims to achieve scalable exact output and regulated output synchronization for discrete-time multi-agent systems in presence of disturbances and measurement noise with known frequencies. Both homogeneous and heterogeneous multi-agent systems are considered, with parts of agents' states accessible in the latter case. The key contribution of this paper is on the distributed protocol that only uses the information of agent models, rather than the communication network information and the agent number, so as to achieve the scalable exact synchronization under disturbances and measurement noise. The validity of the protocol is verified by numerical simulations with arbitrarily chosen number of agents.
\end{abstract}


\section{Introduction}
\label{sec:introduction}

In recent decades, more and more attentions have been drawn to multi-agent systems (MAS). One of the research focuses is the synchronization problem in light of its potential applications, such as automotive vehicle
control, satellites/robots formation, sensor networks, and so forth. Some of the existing work can be found in \cite{bai-arcak-wen,bullobook,kocarev-book, mesbahi-egerstedt,ren-book, saberi-stoorvogel-zhang-sannuti,wu-book} and the references therein.

In the existing work, it can be found that the agent number and the communication network information, such as lower bound and upper bound of the eigenvalues of the associated Laplacian matrix, are needed in the protocol design. It results in \emph{scale fragility} as mentioned in \cite{studli-2017,tegling-bamieh-sandberg,tegling-midelton-seron}. In other words, the stability of the whole MAS might be lost if an agent is added into or dropped from the network, or the connection among agents is changed.

To cope with the issue of scale fragility, we have designed scalable protocol for the synchronization of MAS in various situations, which can be seen in \cite{liu-nojavanzedah-saberi-2022-book}. Our ``scale-free'' protocol design only uses the information of agent models. That is, the communication network information and  the agent number are not needed in the protocol design. In other words, the scalable protocol can be applied to any communication network with a spanning tree.

When the external disturbances affect agents, the almost synchronization of MAS is considered. According to \cite{sab}, the ``almost synchronization'' is usually associated with the concept of almost disturbance decoupling, in which it aims to design a family of protocols to reduce the noise sensitivity as much as possible. So far, scale-free protocol design has been investigated on two kinds of external disturbances: deterministic disturbances subject to finite power and stochastic disturbances under bounded variance. In the former case, $H_\infty$ almost state synchronization is achieved for homogeneous MAS in \cite{liu-saberi-stoorvogel-donya-almost-automatica}, with the use of  properly designed localized information exchange is used. While in the latter case, a scalable $H_2$ almost state synchronization was realized in \cite{liu-saberi-stoorvogel-donya-arxiv-h2almost,liu-saberi-stoorvogel-donya-CCC-12}.

However, if the frequencies of the disturbances and measurement noise are available, exact synchronization of MAS can be obtained. In \cite{zhang-saberi-stoorvogel-acc15-2} and \cite{zhang-saberi-stoorvogel-cdc2016}, the exact output synchronization is achieved for continuous-time heterogeneous MAS with non-introspective agents\footnote{If the agents have access to their own dynamics in addition to relative information from the network, they are called introspective; otherwise, non-introspective.} and under time-varying directed communication network. 

In this paper, we focus on scalable exact output synchronization for discrete-time MAS  in the presence of disturbances and measurement noise with known frequencies. That is, we design scalable linear dynamic protocols to achieve scalable exact output synchronization and scalable exact regulated output synchronizations for homogeneous MAS with non-introspective agents and heterogeneous MAS with introspective agents, respectively. In the homogeneous case, the scalable protocol can be applied to communication network with any agent number as long as the communication network has a spanning tree. While in the heterogeneous case, the scalable protocol can be applied to any communication network with any agent number as long as there exists a path from the exosystem, which generates reference trajectory, to each agent in the communication network.

\subsection*{Notations}
In this paper, $A\T$ is the
conjugate transpose of matrix $A\in \mathbb{R}^{m\times n}$. If $A$ is square and all its eigenvalues are in the open unit circle, the matrix $A$ is Schur stable.  $A\otimes B$ denotes the Kronecker product between matrices $A$ and $B$. And $I_n$ and $0_n$  are identity matrix and zero matrix with dimension $n$ , respectively. The dimension might be neglected in context.

In a MAS, a \emph{weighted graph} $\mathcal{G}$ is utilized to model communication among agents. A triple
$(\mathcal{V}, \mathcal{E}, \mathcal{A})$ is used to define such a  weighted graph $\mathcal{G}$, where $\mathcal{V}=\{1,\ldots, N\}$ for $N$ nodes, $\mathcal{E}$ is a set of
pairs of nodes indicating connections among all nodes, and
$\mathcal{A}=[a_{ij}]\in \mathbb{R}^{N\times N}$ is a weighted adjacency matrix with $a_{ij}\geq0$. Each pair $(j,i)\in \mathcal{E}$ means an \emph{edge} from node $j$ to node $i$  with its weight $a_{ij}>0$. So, $a_{ij}=0$ means no
edge from node $j$ to node $i$. And we assume that $a_{ii}=0$, i.e.,  no self-loop in the graph. A \emph{path} from node $i_1$ to $i_k$ is determined by a
sequence of nodes $\{i_1,\ldots, i_k\}$ with all pairs of nodes
$(i_j, i_{j+1})\in \mathcal{E}$ ($j=1,\ldots, k-1$).  If a graph has a directed spanning tree, there exists a directed path from its root to every other node in the graph \cite{royle-godsil}. The
\emph{weighted in-degree} of node $i$ is given by $d_{\text{in}}(i) = \sum_{j=1}^N\, a_{ij}$.


For a weighted graph $\mathcal{G}$, its associated \emph{Laplacian matrix} $L=[\ell_{ij}]$ is defined as
\[
\ell_{ij}=
\begin{system}{cl}
	\sum_{k=1}^{N} a_{ik}, & i=j,\\
	-a_{ij}, & i\neq j,
\end{system}
\]
Notice that all the eigenvalues of the Laplacian matrix $L$ are in the
closed right half plane with at least one zero eigenvalue associated
with right eigenvector $\textbf{1}$ \cite{royle-godsil}. Moreover, if the graph contains a directed spanning tree, the Laplacian matrix $L$ has just one zero eigenvalue \cite{ren-book}.
%

\begin{definition}\label{def1}
	Let $\mathbb{G}^N$ denote the set of directed graphs of $N$ agents that contain a directed spanning tree.
\end{definition}

\section{Scalable output synchronization of discrete-time homogeneous MAS with known frequency disturbances}
\subsection{Problem formulation} 
The agent dynamics in a discrete-time homogeneous MAS is written in the form of
\begin{equation}\label{homo_sys}
	\begin{system}{cl}
		{x}_i(k+1) &= Ax_i(k) +B u_i(k)+E \omega_i(k),  \\
		y_i(k) &= Cx_i(k)+G\omega_i(k),
	\end{system}
\end{equation}
for $i=1,\ldots, N$,  where $x_i(k)\in\mathbb{R}^{n}$, $u_i(k)\in\mathbb{R}^{m}$ and $y_i(k)\in\mathbb{R}^{p}$ are agent $i$'s state,
input, output. The external disturbance $\omega_i$ has know frequencies, and can be represented by an exosystem
\begin{equation}\label{hekf-distsys}
	{\omega}_{i}(k+1)=S\omega_i(k),
\end{equation}
with its initial conditions arbitrarily chosen, and the states
$\omega_i(k)\in \R^{n_\omega}$.  Then,  the process disturbances with known frequencies is modeled as $E\omega_i$ and the measurement noise with known frequencies is modeled as $G\omega_i$, respectively.

\begin{figure*}[h]
	\includegraphics[width=13cm]{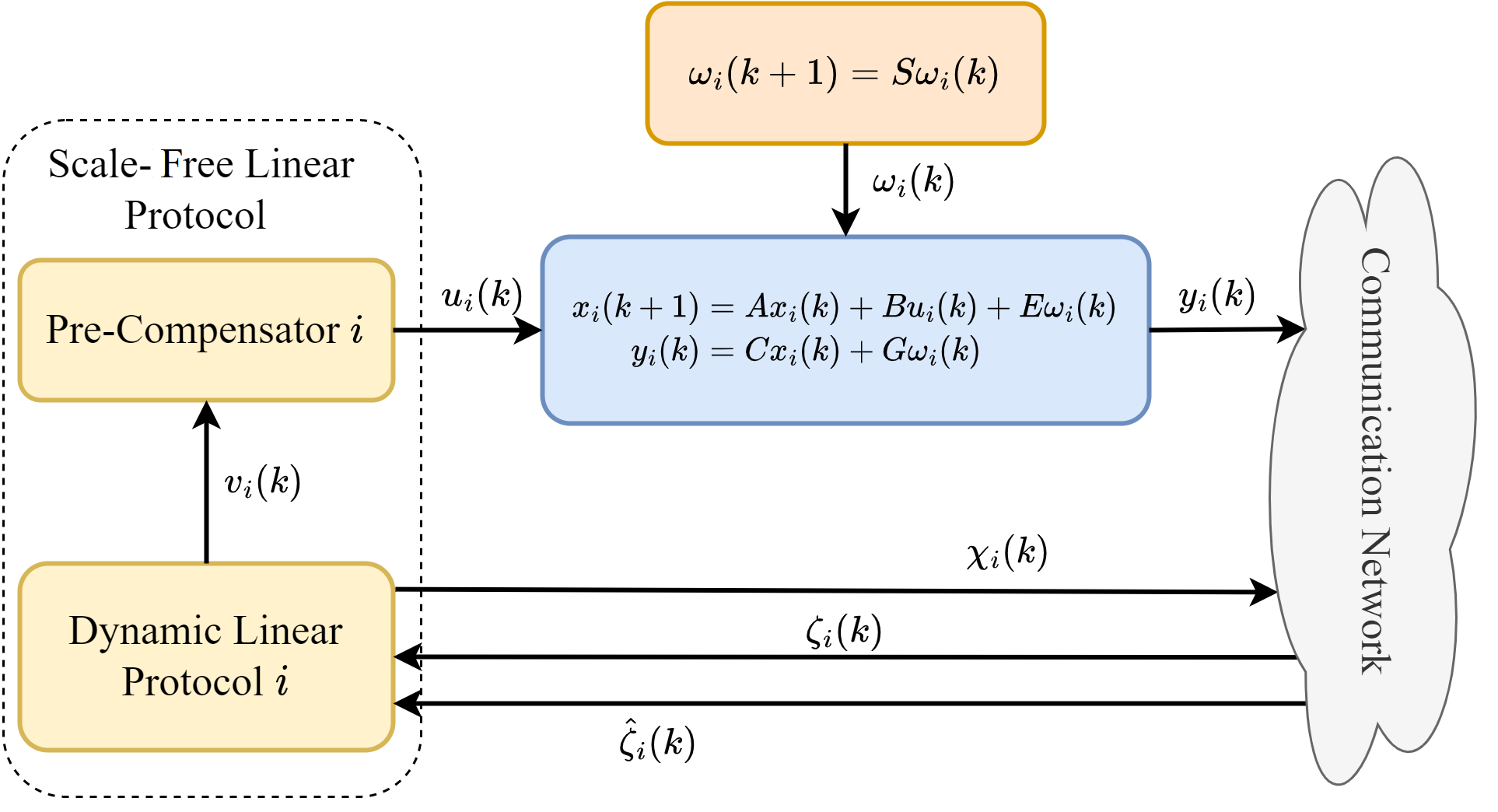}
	\centering
	\caption{Architecture of the scalable linear protocol for homogeneous MAS}\label{output-syn}
\end{figure*}

The scalable protocol only utilizes the localized information, which consists of two types of information. One is the relative output information from communication network, which is represented by 
\begin{equation}\label{zeta1}
	\zeta_i(k)=\frac{1}{1+\bar{d}_{\text{in}}(i)}\sum_{j=1}^{N}a_{ij}(y_i(k)-y_j(k)),
\end{equation}
where $a_{ij}>0$, $a_{ii}=0$, and $\bar{d}_{\text{in}}(i)$ is the \emph{upper bound} on weighted in-degree of agent $i$, i.e., $d_{\text{in}}(i)$ for $i=1,\cdots,N$. Obviously, the $\bar{d}_{\text{in}}(i)$ is still a local information. And $\zeta_i(k)$ can be rewritten as
\begin{equation}\label{zeta}
	\zeta_i(k)=\sum_{j=1,j\neq i}^N d_{ij}(y_i(k)-y_j(k)),
\end{equation}
where $d_{ij}\geq 0$, and we choose
$d_{ii}=1-\sum_{j=1,j\neq i}^Nd_{ij}$ such that $\sum_{j=1}^Nd_{ij}=1$
with $i,j\in\{1,\ldots,N\}$. The matrix $D=[d_{ij}]_{N\times N}$ is then a so-called, row
stochastic matrix. Let $D_{\text{in}}=\diag\{\bar{d}_{\text{in}}(1),\bar{d}_{\text{in}}(2),\cdots,\bar{d}_{\text{in}}(N) \}$. Then the
relationship between the row stochastic matrix $D$ and the Laplacian
matrix $L$ is
\begin{equation*}
	(I+D_{\text{in}})^{-1}L=I-D.
\end{equation*}

And the other is a localized information exchange among the agents, which is represented by
\begin{equation}\label{etahat}
	\hat{\zeta}_i(k)=\sum_{j=1}^{N}d_{ij}(\eta_i(k)-\eta_j(k))
\end{equation}
where $\eta_i(k)$ is part of the state of agent $i$'s protocol.

Next, we formulate the problem of scalable output synchronization as follows.
\begin{problem}\label{prob_reg_x}
	For a discrete-time MAS modeled by \eqref{homo_sys} and \eqref{zeta} in the presence of disturbances generated by exosystem \eqref{hekf-distsys},
	the \textbf{scalable output synchronization} is to design, if possible, a linear dynamic protocol, only using the agent's model, i.e. $(C, A, B, E, G, S)$, of the form:
	\begin{equation}\label{out_reg_dyn}
		\begin{system}{cl}
			{x}_{i,c}(k+1)&=A_{ c}x_{i,c}(k)+ B_{c}{\zeta}_i(k)+ C_{c}\hat{\zeta}_i(k),\\
			u_i(k)&=F_{c}x_{i,c}(k),
		\end{system}
	\end{equation}
	where $x_{i,c}(k)\in \mathbb{R}^{n_c}$ and $\hat{\zeta}_i(k)$ is given by \eqref{etahat} with $\eta_i(k)=M_c x_{i,c}(k)$ and $M_c$ is a constant matrix,
	such that the output synchronization among agents is achieved, i.e., 
	\begin{equation}\label{synch_state}
		\lim\limits_{k\to\infty}(y_i(k)-y_j(k))=0,
	\end{equation} 
	for any $N$, any fixed graph $\mathscr{G}\in \mathbb{G}^N$, all $i,j \in \{1,\hdots, N\}$, and any process disturbances and measurement noise generated by exosystem \eqref{hekf-distsys}.
\end{problem}

\begin{remark}
	It is worth noticing that agents' states or outputs are not involved in the above protocol \eqref{out_reg_dyn},  which confirms that agents are non-introspective in this section.
\end{remark}

To ensure the achievability of scalable output synchronization, we assume that
\begin{assumption}\label{ass3}\	
	
	\begin{enumerate}
		\item[A1.1] All the eigenvalues of $A$ are in the closed unit circle, i.e., agents are weakly unstable.
		\item[A1.2] All the eigenvalues of $S$ are on the unit circle.
		\item[A1.3] $(A,B)$ is stabilizable.
		\item[A1.4] $\left(\begin{pmatrix}
			C&-G
		\end{pmatrix}, \begin{pmatrix}
			A&-E\\0&S
		\end{pmatrix}
		\right)$ is detectable.
		\item[A1.5] $(C,A,B)$ is right-invertible.
		\item[A1.6] $(C,A,B)$ has no invariant zeros on or outside the unit circle, and the invariant zeros of $(C,A,B)$ do not intersect with eigenvalues of $S$.		
	\end{enumerate}
\end{assumption}

\begin{remark}
	Assumptions A1.5 and A1.6 guarantee the existence of a solution to the so-called regulation equations. These conditions can be weakened (see \cite[Chapter 2]{saberi-stoorvogel-sannutia}). Meanwhile, Assumption A1.4 can be reduced to $(C,A)$ detectable, using the technique from \cite[Chapter 2]{saberi-stoorvogel-sannutia}.
\end{remark}

\subsection{Protocol design in homogeneous MAS}

The scale-free linear protocol design architecture is
shown in Figure \ref{output-syn}. It is fulfilled 
in three steps as follows. On Step $1$, a pre-compensator is designed to include the disturbance mode and on Step $2$, a scalable protocol is designed for the interconnected system of the agent and the pre-compensator on Step $1$. Then, on Step $3$, the pre-compensator on Step $1$ and the scalable protocol on Step $2$ are combined into the over-all protocol presented in \eqref{out_reg_dyn}.\\

\noindent{\textbf{Step 1:  Designing a pre-compensator for internal model principle}  }

The regulation equations
\begin{equation}\label{regeq}
	\begin{system}{l}
		\Pi{S}=A\Pi+B\Gamma+E, \\		
		C\Pi+G=0,
	\end{system}
\end{equation}
are solvable for $\Gamma\in \R^{m\times n_{\omega}}$ and $\Pi\in \R^{n\times n_{\omega}}$ when Assumptions A1.5 and A1.6 are satisfied (see \cite[Chapter 2]{saberi-stoorvogel-sannutia}). Then, we design
a pre-compensator as
\begin{equation}\label{compensator1a}
	\begin{system}{l}
		{p}_i(k+1)=Sp_i(k)+B_{p}v_i(k),\\
		{u}_i(k)=\Gamma p_i(k)+D_pv_i(k),
	\end{system}
\end{equation}
for $i=1,\hdots, N$, where $p_i(k)\in \R^{n_{\omega }}$, $B_{p}$ and $D_p$ are chosen based on \cite{liu-lin-chen-tac-2009} such that pre-compensator \eqref{compensator1a} is invertible and of minimum-phase, which ensures that the cascade system of \eqref{homo_sys} and \eqref{compensator1a} is stablilizable with respect to $v_i(k)$. 


%

Define $
\tilde{x}_i(k)=\begin{pmatrix}
	x_i(k)\T&p_i(k)\T
\end{pmatrix}\T$, and
\begin{align*}
	&\tilde{A}=\begin{pmatrix}
		A&B\Gamma\\0&S
	\end{pmatrix}, \tilde{B}=\begin{pmatrix}
		BD_p\\B_{p}
	\end{pmatrix},\tilde{E}=\begin{pmatrix}
		E\\0
	\end{pmatrix}, \tilde{C}=\begin{pmatrix}
		C&0
	\end{pmatrix}. 
\end{align*} Then, we can rewrite the cascade system of pre-compensator \eqref{compensator1a} and agent system \eqref{homo_sys} as,
\begin{equation}\label{hekf-system-ana}
	\begin{system}{cl}
		{\tilde{x}}_i(k+1) &=\tilde{A}\tilde{x}_i(k)+\tilde{B}v_i(k)+\tilde{E}\omega_i(k),\\
		y_i(k)&=\tilde{C}\tilde{x}_i(k)+{G}\omega_i(k),
	\end{system}
\end{equation}
in which $\tilde{x}_{i}\in\R^{\tilde{n}}$, $v_{i}\in\R^{p},\,y_{i}\in\R^{p}$ are states, inputs and outputs of the cascade system. Moreover, under Assumption A1.4, ($\tilde{A}$, $\tilde{B}$) is stabilizable and ($\tilde{C}, \tilde{A}$) is detectable (see the proof in Appendices \ref{app01} and \ref{app1}).

It is obvious that
$(\tilde{A}, \tilde{C}, \tilde{E}, G)$ satisfies
\begin{equation}\label{aregeq}
	\begin{system}{ll}
		&\tilde{\Pi}
		S=\tilde{A}\tilde{\Pi}+\tilde{E},\\
		&\tilde{C}\tilde{\Pi}+G=0,
	\end{system}		
\end{equation}
with
$
\tilde{\Pi}=\begin{pmatrix}
	\Pi\\I
\end{pmatrix}$.

Next, we define a new variable $\bar{x}_i(k)=\tilde{x}_i(k)-\tilde{\Pi} \omega_{i}(k)$, which leads to a new dynamics with the same output:
\begin{align*}
	{\bar{x}}_i(k+1)&=\tilde{A}\tilde{x}_i(k)+\tilde{B}_iv_i(k)+\tilde{E}\omega_i(k)-\tilde{\Pi} S\omega_i(k)\\
	&=\tilde{A}\tilde{x}_i(k)+\tilde{B}_iv_i(k)+\tilde{E}\omega_i(k)-(\tilde{A}\tilde{\Pi}+\tilde{E})\omega_i(k)\\
	&=\tilde{A}\bar{x}_i(k)+\tilde{B}_iv_i(k),
\end{align*}
and 
\begin{align*}
	{y}_i(k)&=\tilde{C}\tilde{x}_i(k)+{G}\omega_i(k)\\
	&=\tilde{C}\bar{x}_i(k)+G\omega_i(k)+\tilde{C}\tilde{\Pi}\omega_i(k)\\
	&=\tilde{C}\bar{x}_i(k).
\end{align*}
Note that disturbances are canceled out in the above dynamics, which is rewritten as follows, 
\begin{equation}\label{hekf-systema}
	\begin{system}{cl}
		{\bar{x}}_i(k+1) &=\tilde{A}\bar{x}_i(k)+\tilde{B}v_i(k),\\
		{y}_i(k)&=\tilde{C}\bar{x}_i(k),
	\end{system}
\end{equation}
where $\bar{x}_{i}(k)\in\R^{\tilde{n}}$, $v_{i}(k)\in\R^{p}$.\\

%
	%
	%

		\noindent\textbf{Step 2: Designing a scalable protocol for the cascade system $\bar{x}_i(k)$}
		
		Matrices $H$ and $K$ are chosen properly to ensure $\tilde{A}-H\tilde{C}$ and $\tilde{A}-\tilde{B}K$ are Schur stable. Then, we design the scalable protocol as follows,
		\begin{equation}\label{pscp1}
			\begin{system}{cl}
				{\hat{x}}_i(k+1)&=\tilde{A}\hat{x}_i(k)-\tilde{B}K\hat{\zeta}_i(k)+H({\zeta}_i(k)-\tilde{C}\hat{x}_i(k)),\\
				{\chi}_i(k+1)&=\tilde{A}\chi_i(k)+\tilde{B}v_i(k)+\hat{x}_i(k)-\hat{\zeta}_i(k),\\
				v_i(k)&=-K\chi_i(k),
			\end{system}
		\end{equation}
		where 
		\begin{equation}\label{hatzetahete}
			\hat{\zeta}_i(k)=\sum_{j=1}^{N}d_{ij}(\chi_i(k)-\chi_j(k)),
		\end{equation} 
		with $\chi_i(k)$ being the internal states in \eqref{etahat}, i.e., $\eta_i(k)=\chi_i(k)$. \\

		\noindent\textbf{Step 3: Obtaining an overall scalable protocol for each agent}
		
		The overall scalable protocol for agent $i$ is a combination of pre-compensator \eqref{compensator1a} on Step $1$ and the scalable protocol \eqref{pscp1} on Step $2$, i.e.,
		\begin{equation}\label{pscp2final}
			\begin{system}{cl}
				{p}_i(k+1)&=Sp_i(k)-B_{p}K\chi_i(k),\\
				{\hat{x}}_i(k+1)&=\tilde{A}\hat{x}_i(k)-\tilde{B}K\hat{\zeta}_i(k)+H({\zeta}_i(k)-\tilde{C}\hat{x}_i(k)),\\
				{\chi}_i(k+1)&=\tilde{A}\chi_i(k)-\tilde{B}K\chi_i(k)+\hat{x}_i(k)-\hat{\zeta}_i(k),\\
				u_i(k)&=\Gamma p_i(k)-D_pK\chi_i(k).
			\end{system}
		\end{equation}
		
		Now the main result in this section can be stated in the following theorem.
		\begin{theorem}\label{thm_f1}
			Consider a discrete-time MAS modeled by \eqref{homo_sys} and \eqref{zeta} in the presence of disturbances generated by \eqref{hekf-distsys}. Let Assumption \ref{ass3} be satisfied. 
			
			Then, the scalable output synchronization problem defined in Problem \ref{prob_reg_x} is solvable. Specifically, the protocol \eqref{pscp2final} achieves output synchronization \eqref{synch_state} among agents for any $N$, any fixed graph
			$\mathscr{G}\in\mathbb{G}^N$, and any process disturbances and measurement noise generated by exosystem \eqref{hekf-distsys}. 
		\end{theorem} 
		
		\begin{proof}[The proof of Theorem \ref{thm_f1}]	
			Define $\rho_i(k)=\bar{x}_i(k)-\bar{x}_N(k)$, $\hat{\rho}_i(k)=\hat{{x}}_i(k)-\hat{{x}}_N(k)$, $\tilde{y}_i(k)={y}_i(k)-{y}_N(k)$, and $\tilde{\chi}_i(k)=\chi_i(k)-\chi_N(k)$. Then, we have
			\begin{equation*}
				\begin{system}{rl}
					{\rho}_i(k+1) &=\tilde{A}\rho_i(k)-\tilde{B}K\tilde{\chi}_i(k),\\
					\tilde{y}_i(k)&=\tilde{C}\rho_i(k),\\
					{\hat{\rho}}_i(k+1)&=\tilde{A}\hat{\rho}_i(k)-\tilde{B}K\sum_{j=1}^{N-1}\bar{d}_{ij}\tilde{\chi}_j(k)+H\tilde{C}(\sum_{j=1}^{N-1}\bar{d}_{ij}\tilde{y}_j(k)-\hat{\rho}_i(k))\\
					{\tilde{\chi}}_i(k+1)&=\tilde{A}\tilde{\chi}_i(k)-\tilde{B}K\tilde{\chi}_i(k)+\tilde{A}\hat{\rho}_i(k)-\sum_{j=1}^{N-1}\bar{d}_{ij}\tilde{A}\tilde{\chi}_j(k),
				\end{system}
			\end{equation*}	
			where $\bar{d}_{ij}=d_{ij}-d_{iN}$.

			Next, by defining
			\begin{align*}
				\rho(k)&=\begin{pmatrix}
					\rho_1(k)\\ \vdots\\ \rho_{N-1}(k)
				\end{pmatrix},\hat{\rho}(k)=\begin{pmatrix}
					\hat{\rho}_1(k)\\ \vdots\\ \hat{\rho}_{N-1}(k)
				\end{pmatrix},\\
				\tilde{\chi}(k)&=\begin{pmatrix}
					\tilde{\chi}_1(k)\\ \vdots\\ \tilde{\chi}_{N-1}(k)
				\end{pmatrix},\tilde{y}(k)=\begin{pmatrix}
					\tilde{y}_1(k)\\ \vdots\\\tilde{y}_{N-1}(k)\end{pmatrix},
			\end{align*}
			the overall disagreement system can be written as follows,
			\begin{equation}
				\begin{system}{cl}
					{\rho}(k+1)&=(I_{N-1}\otimes\tilde{A})\rho(k)-(I_{N-1}\otimes \tilde{B}K)\tilde{\chi}(k)\\
					{\hat{\rho}}(k+1)&=( I_{N-1}\otimes(\tilde{A}-H\tilde{C} ))\hat{\rho}(k)-((I_{N-1}-\bar{D})\otimes \tilde{B}K)\tilde{\chi}(k)\\
					&\hspace{6cm}+((I_{N-1}-\bar{D})\otimes H\tilde{C})\rho(k)\\
					{\tilde{\chi}}(k+1)&=(\bar{D}\otimes \tilde{A})\tilde{\chi}(k) -(I_{N-1}\otimes\tilde{B}K)\tilde{\chi}(k)+(I_{N-1}\otimes \tilde{A})\hat{\rho}(k)\\
					\tilde{y}(k)&=(I_{N-1}\otimes \tilde{C})\rho(k)
				\end{system}
			\end{equation}
			with $\bar{D}=[\bar{d}_{ij}]_{(N-1)\times (N-1)}=[d_{ij}-d_{iN}]_{(N-1)\times (N-1)}$. According to \cite[Lemma 1]{donya-liu-saberi-stoorvogel-arxiv-dis}, we have all the eigenvalues of $\bar{D}$ are inside the open unit circle.

			Now let $\delta(k)=\rho(k)-\tilde{\chi}(k)$ and $\bar{\delta}(k)=((I_{N-1}-\bar{D})\otimes I)\rho(k)-\hat{\rho}(k)$, we can obtain  
			\begin{equation}\label{newsystem2}
				\begin{system}{cl}
					{\rho}(k+1)&=(I_{N-1}\otimes( \tilde{A}-\tilde{B}K))\rho(k)+(I_{N-1}\otimes \tilde{B}K )\delta(k),\\
					{\delta}(k+1)&=(\bar{D}\otimes \tilde{A})\delta(k)+(I_{N-1}\otimes\tilde{A})\bar{\delta}(k),\\
					{\bar{\delta}}(k+1)&=(I_{N-1}\otimes(\tilde{A}-H\tilde{C}))\bar{\delta}(k).
				\end{system}
			\end{equation}
			Next, we will prove that the above system \eqref{newsystem2} is Schur stable, i.e., $\lim_{k\to\infty}\rho_i(k)\to 0$, which leads to $\lim_{k\to\infty}\tilde{y}_i(k)\to 0$. This implies that $y_i(k)-y_N(k)\to 0$ as $k\to \infty$ for all $i=1,\ldots,N-1$. This proves the achievement of scable output synchronization.
			
			
			According to \cite{liu-saberi-stoorvogel-IJRNC-2022}, we have $\bar{D}\otimes \tilde{A}$ is Schur stable since all eigenvalues of $\tilde{A}$ are in closed unit circle and all eigenvalues of $\bar{D}$ are in the open unit circle.
			Moreover, $\tilde{A}-\tilde{B}K$ and $\tilde{A}-H\tilde{C}$ are Schur stable for properly chosen matrices $K$ and $H$. Therefore, the system \eqref{newsystem3} is asymptotically stable, i.e., 
			\[
			\lim_{k\to\infty}\rho_i(k)\to 0.
			\]
			Meanwhile, since $\tilde{y}_i(k)=\tilde{C}\rho_i(k)$, $\lim_{k\to\infty}\tilde{y}_i(k)\to 0$, i.e. $\lim_{k\to\infty}y_i(k)\to y_j(k)$, which proves the result.
		\end{proof}
		
		\begin{remark}
			Notice that state synchronization of the cascade system \eqref{hekf-systema} is achieved in the above proof, i.e., $\bar{x}_i(k)\to 0$, which means 
			\[
			x_i(k)-x_j(k)\to \Pi(\omega_i(k)-\omega_j(k)).
			\]	
			Since disturbances $\omega_i(k)$ might not be identical for $i=1,\cdots, N$, state synchronization can not be achieved among the original agents. However, obviously, in the absence of disturbances, the protocol \eqref{pscp2final} achieve state synchronization.
			
		\end{remark}
		
		\section{Scalable regulated output synchronization of heterogeneous discrete-time MAS with known frequencies disturbances}
		\subsection{Problem formation}
		The agent dynamics in a discrete-time heterogeneous MAS is written in the form of
		\begin{equation}\label{hete_sys}
			\begin{system}{cl}
				{x}_i(k+1) &= A_ix_i(k) +B_i u_i(k)+E_i \omega_i(k),  \\
				y_i(k) &= C_ix_i(k)+G_i\omega_i(k),
			\end{system}
		\end{equation}
		for $i=1,\ldots, N$, where $x_i(k)\in\mathbb{R}^{n_i}$, $u_i(k)\in\mathbb{R}^{m_i}$, and $y_i(k)\in\mathbb{R}^{p}$ represent agent $i$'s state,
		input, and output, respectively. We assume all agents are introspective, i.e., each agent has access to part of agent's states defined as,
		\begin{equation}\label{local}
			z_i(k)=C_i^mx_i(k),
		\end{equation}
		where $z_i(k)\in \mathbb{R}^{z_i}$. Now the external disturbance $\omega_i$ is generated by the following exosystem:
		\begin{equation}\label{hekf-distsys2}
			{\omega}_{i}(k+1)=S_i\omega_i(k),
		\end{equation}
		with any arbitrarily chosen initial conditions and $\omega_i(k)\in \R^{n_{\omega_i}}$. Again, $E_i\omega_i$ and $G_i\omega_i$
		stands for the process disturbances and the measurement noise with different known frequencies, respectively. 
		
		
		Similarly, our goal in this section is to design a linear protocol to achieve regulated output synchronization, i.e., 
		\begin{equation}\label{reg_synch_state}
			\lim\limits_{k\to\infty}(y_i(k)-y_r(k))=0
		\end{equation}
		for any $i\in\{1,\ldots,N\}$, where $y_r(k)$ is a priori given trajectory generated by an exosystem
		\begin{equation}\label{exo}
			\begin{system}{cl}
				{x}_r(k+1)&=A_r x_r(k), \\
				y_r(k)&=C_r x_r(k),
			\end{system}\quad x_r(0)=x_{r0},
		\end{equation}
		where $x_r(k) \in\mathbb{R}^{r}$ and $y_r(k)\in\mathbb{R}^p$. And the following assumptions should be satisfied:
		\begin{assumption}\label{ass-exo}\ 
			
			\begin{enumerate}
				\item[A2.1]  All the eigenvalues of $A_r$ are on the unit circle.
				\item[A2.2] $(C_r, A_r)$ is observable.
				
			\end{enumerate}
		\end{assumption}

		It is well known that, to enable the regulation among agents, some of the agents, defined as a set of agents $\mathscr{C}$,  should have access to the output of the exosystem. Moreover, there exists paths from all the other agents to agents in the set. In other words, agents in the set obtain extra quantity from the exosystem as,
		\begin{equation}\label{q_psi}
			\Psi_i(k)=\iota_i(y_i(k)-y_r(k)), \qquad \iota_i=\begin{system}{cl}
				1, \quad i\in \mathscr{C},\\
				0, \quad i\notin \mathscr{C}.
			\end{system}
		\end{equation}
		And, we define a set of graphs specially for scalable regulated output synchronization as follows.
		\begin{definition}\label{def_rootset}
			Given a node set $\mathscr{C}$, $\mathbb{G}_{\mathscr{C}}^N$ denotes a set of all graphs with $N$ nodes containing the node set $\mathscr{C}$, such that every node of the graph $\mathcal{G}\in\mathbb{G}_\mathscr{C}^N$ is a member of a directed tree
			that has its root included in the node set $\mathscr{C}$. The node set $\mathscr{C}$ is referred to as root set.
		\end{definition}
		
		\begin{remark}
			Notice that Definition \ref{def_rootset} does not need the existence of a directed spanning tree in the graph .
		\end{remark}			
		
		\begin{figure*}[h]
			\includegraphics[width=13cm]{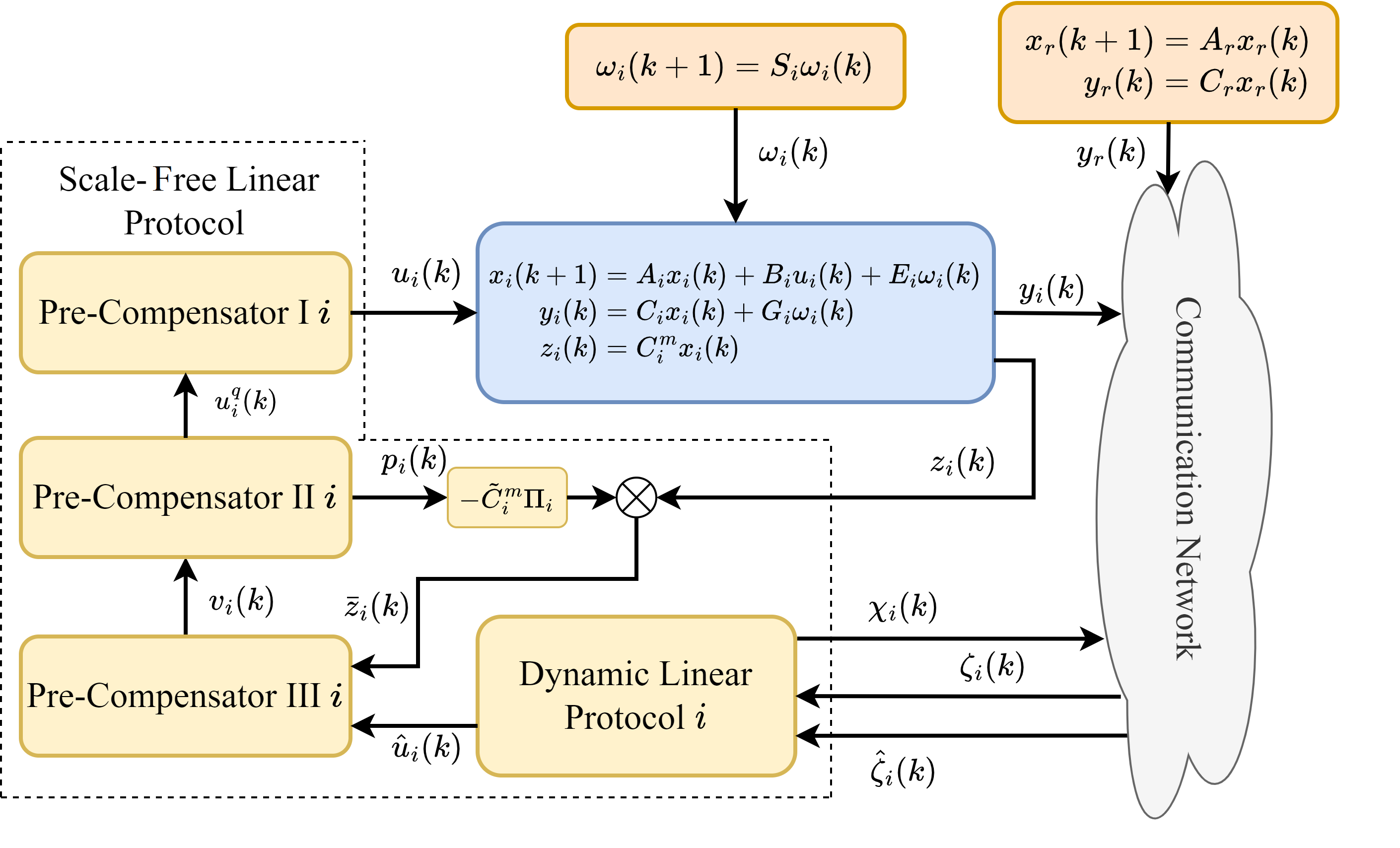}
			\centering
			\caption{Architecture of the scalable linear protocol for heterogeneous MAS}\label{reg-output-syn}
		\end{figure*}
		
		When integrating the extra quantity \eqref{q_psi} into \eqref{zeta1} with a new scaling part, we obtain the updated relative information from communication network, i.e.,
		\begin{equation}\label{zetabar}
			\tilde{\zeta}_i(k)=\frac{1}{2+\bar{d}_{\text{in}}(i)}\sum_{j=1}^{N}a_{ij}(y_i(k)-y_j(k))+\iota_i(y_i(k)-y_r(k)).
		\end{equation}
		In terms of the coefficients of a so-called expanded Laplacian matrix $\tilde{L}=L+\diag\{\iota_i\}=[\tilde{\ell}_{ij}]_{N \times N}$, $\tilde{\zeta}_i(k)$ can be rewritten as
		\begin{equation}\label{zetabar2}
			\tilde{\zeta}_i(k)=\frac{1}{2+\bar{d}_{\text{in}}(i)}\sum_{j=1}^{N}\tilde{\ell}_{ij}(y_j(k)-y_r(k)).
		\end{equation}
		%
		%
		%

		For any graph $\mathcal{G}\in \mathbb{G}_\mathscr{C}^N$, with the associated expanded Laplacian matrix $\tilde{L}$, we define 
		\begin{equation*}
			\tilde{D}=I_N-(2I_N+D_{\text{in}})^{-1}{\tilde{L}}=[\tilde{d}_{ij}]_{N\times N}.
		\end{equation*}
		According to \cite[Lemma 1]{liu2018regulated}, we have all eigenvalues of $\tilde{D}$ are in the open unit circle.

		Meanwhile, a localized information exchange among the agents is represented by	
		\begin{equation}\label{eqa1}
			\hat{\zeta}_i(k)=\frac{1}{2+\bar{d}_{\text{in}}(i)}\sum_{j=1}^Na_{ij}(\xi_i(k)-\xi_j(k))
		\end{equation}
		with $\xi_j(k)$ from agent $j$'s protocol and defined in the following sections.
		
		
		Now we formulate the problem of \textbf{scalable} regulated output synchronization as follows.
		\begin{problem}\label{prob_reg_xheter}
			For a discrete-time heterogeneous MAS modeled by \eqref{hete_sys}, \eqref{local}, and \eqref{zetabar} in the presence of disturbances generated by \eqref{hekf-distsys2}, and reference trajectory generated by \eqref{exo}, the \textbf{scalable regulated output synchronization} is to find, if possible, a linear dynamic protocol, using the only knowledge of agent models, i.e. $(C_i, A_i, B_i, E_i,G_i,C_i^m,S_i,A_r, C_r)$, of the form:
			\begin{equation}\label{out_reg_dyn-hete}
				\begin{system}{cl}
					{x}_{i,c}(k+1)&=A_{i,c}x_{i,c}(k)+ B_{i,c}\tilde{\zeta}_i(k)+ C_{i, c}\hat{\zeta}_i(k)+D_{i,c}z_i(k),\\
					u_i(k)&=E_{c}x_{i,c}(k)+ F_{i,c}\tilde{\zeta}_i(k)+ G_{i, c}\hat{\zeta}_i(k)+H_{i,c}z_i(k),
				\end{system}
			\end{equation}
			where $x_{i,c}(k)\in \mathbb{R}^{n_{c_i}}$ and $\hat{\zeta}_i(k)$ is given by \eqref{eqa1} with $\xi_i(k)=M_c x_{i,c}(k)$ and $M_c$ is a constant matrix,
			such that the regulated output synchronization \eqref{reg_synch_state} among agents is achieved, i.e., 
			\begin{equation*}
				\lim\limits_{k\to\infty}(y_i(k)-y_r(k))=0
			\end{equation*}
			for any $N$, any fixed graph $\mathscr{G}\in \mathbb{G}_\mathscr{C}^N$, any process disturbances and measurement noise generated by \eqref{hekf-distsys2}, and any reference trajectory given by \eqref{exo}.
		\end{problem}

		To solve the above scalable regulated output synchronization problem,  for agent $i \in \{1,..., N\}$ we assume that
		\begin{assumption}\label{ass1}\ 
			
			\begin{enumerate}
				\item[A3.1] $(A_i,B_i)$ is stabilizable.
				\item[A3.2] $\left(\begin{pmatrix}
					C_i&-G_i
				\end{pmatrix}, \begin{pmatrix}
					A_i&-E_i\\0&S_i
				\end{pmatrix}
				\right)$ is detectable.
				\item[A3.3] $(C_i,A_i,B_i)$ is right-invertible.
				\item[A3.4] $\left(\begin{pmatrix}
					C_i^m&0
				\end{pmatrix}, \begin{pmatrix}
					A_i&-E_i\\0&S_i
				\end{pmatrix}
				\right)$ is detectable.
				\item[A3.5]  $(C_i,A_i,B_i)$ has no invariant zeros on or outside of unit circle that coincide with the eigenvalues of $S_i$ .
			\end{enumerate}
		\end{assumption}
		
		\begin{remark}
			Assumptions A3.2 and A3.4 can be reduced to that, $(C_i, A_i)$ is detectable and the technical condition
			\[
			\ker\begin{pmatrix}
				\lambda I-A_i&E_i\\0&\lambda I-S_i\\C_i&-G
			\end{pmatrix}=\ker\begin{pmatrix}
				\lambda I-A_i&E_i\\0&\lambda I-S_i\\C_i&-G\\
				C_i^m&0
			\end{pmatrix}
			\]
			for all $\lambda\in \mathbb{C}$, which is actually a necessary condition (see \cite[Chapter 2]{saberi-stoorvogel-sannutia}).
			Moreover, conditions on the eigenvalues of $A_i$ and $S_i$ are not required in this scale-free protocol design for heterogeneous MAS.
		\end{remark}
		
		\subsection{Protocol design in heterogeneous MAS}\label{hete-pro-desi}
		
		The scale-free linear protocol design architecture is
		shown in Figure \ref{reg-output-syn}. I can be realized in six steps. On Step $1$, a pre-compensator is designed to make the agent model invertible. In Step $2$, another pre-compensator is designed to include the mode of the disturbance exosystem into the agent model. While Step $3$ shapes the reference exosystem into a specific form. And Step $4$ almost homogenize the cascaded system of the agent model and two pre-compensators from the above steps. Then, on Step $5$, a scalable protocol is designed for the almost homogenized system from Step $4$. Finally, we integrate all pre-compensators and the scalable protocol and obtain an overall scalable protocol for the original agent.\\
		
			
			\noindent{\textbf{Step 1:  Designing pre-compensator I to make the agent models invertible}  }
			
			According to the design process in \cite[Section 3.B]{ss} and \cite[Chapter 23]{saberi-stoorvogel-zhang-sannuti}, a discrete-time pre-compensator that is asymptotically stable can be developed as
			\begin{equation}\label{precomz}
				\begin{system}{rl}
					{q}_i(k+1)=&A_{q,i}q_i(k)+B_{q,i}u_i^q(k),\\
					u_i(k)=&C_{q,i}q_i(k)+D_{q,i}u_i^q(k),
				\end{system}	
			\end{equation}
			where $q_i(k)\in \R^{q_i}$ and $u_i^q(k)\in \R^{m_{q_i}}$. Then, the cascade system of this pre-compensator and the agent system \eqref{hete_sys} is invertible.
			Define
			\begin{align*}	
				&\eta_i=\begin{pmatrix}
					x_i\\
					q_i
				\end{pmatrix}, \quad\tilde{A}_{q,i}=\begin{pmatrix}
					A_i&B_iC_{q,i}\\0&A_{q,i}
				\end{pmatrix}, \quad
				\tilde{B}_{q,i}=\begin{pmatrix}
					B_iD_{q,i}\\B_{q,i}
				\end{pmatrix}, \\
				&\tilde{E}_{q,i}=\begin{pmatrix}
					E_i\\0
				\end{pmatrix}, \quad
				\tilde{C}_{q,i}=\begin{pmatrix}
					C_i&0
				\end{pmatrix},\quad
				\tilde{C}_{q,i}^m=\begin{pmatrix}
					C_i^m&0
				\end{pmatrix}.
			\end{align*}
			Then, this cascade system can be written as
			\begin{equation}\label{newsys}
				\begin{system}{rl}
					{\eta}_i(k+1)&=\tilde{A}_{q,i}{\eta}_i(k)+\tilde{B}_{q,i}u_i^q(k)+\tilde{E}_{q,i}\omega_{i}(k),\\
					y_i(k)&=\tilde{C}_{q,i}\eta_i(k)+G_i\omega_{i}(k),\\
					z_i(k)&=\tilde{C}_{q,i}^m\eta_i(k),
				\end{system}
			\end{equation}
			which has the following properties:
			\begin{itemize}
				\item ($\tilde{C}_{q,i},\tilde{A}_{q,i}, \tilde{B}_{q,i}$) is stabilizable, detectable, and invertible.
				\item ($\tilde{C}_{q,i}^m, \tilde{A}_{q,i}$) is detectable
				\item Poles of system \eqref{newsys} are the poles of agent \eqref{hete_sys} plus the poles of pre-compensator
				(i.e., the eigenvalues of $A_{q,i}$ are in open left-half plane).
				\item Infinite zero structure of ($\tilde{C}_{q,i},\tilde{A}_{q,i}, \tilde{B}_{q,i}$) is the same as the infinite zero structure of the system
				(${C}_{i},{A}_{i}, {B}_{i}$).
				\item Invariant zeros of ($\tilde{C}_{q,i},\tilde{A}_{q,i}, \tilde{B}_{q,i}$) are the invariant zeros of the system (${C}_{i},{A}_{i}, {B}_{i}$) plus some additional invariant zeros that can be randomly placed in the open
				unit circle.
			\end{itemize}
			\vspace{3mm}

			\noindent{\textbf{Step 2:  Designing pre-compensator II for internal model principle}  }
			
			Similar to Step $1$ in the homogeneous case, the regulation equations:
			\begin{equation}\label{regul-heter}
				\begin{system}{l}		
					\Pi_{i}{S}_i=\tilde{A}_{q,i}\Pi_{i}+\tilde{B}_{q,i}\Gamma_{i}+\tilde{E}_{q,i}, \\
					\tilde{C}_{q,i}\Pi_{i}+G_i=0,
				\end{system}
			\end{equation}
			have solutions $\Gamma_{i}\in \R^{m_{i}\times n_{\omega i}}$, $\Pi_{i}\in \R^{(n_{i}+n_{q_i})\times (n_{r}+n_{\omega i})}$.
			Then, we design a pre-compensator 
			\begin{equation}\label{compensator1a2}
				\begin{system}{l}
					{p}_i(k+1)=S_{i}p_i(k)+B_{p,i}v_i(k),\\
					{u}_i^q(k)=\Gamma_{i}p_i(k)+D_{p,i}v_i(k),
				\end{system}
			\end{equation}
			where $p_i(k)\in \R^{n_{\omega i}}$, $B_{p,i}$ and $D_{p,i}$ are selected accoding to \cite{liu-lin-chen-tac-2009} to make sure that this pre-compensator is invertible and of minimum-phase. This also guarantees that the cascade system of this pre-compensator and the agent system \eqref{hete_sys} is stablizable with respect to $v_i(k)$.
		
		Now define $\tilde{x}_{i}(k)=\begin{pmatrix}
			\eta_i(k)\T&p_i(k)\T
		\end{pmatrix}\T$, and
		\begin{equation*}
			\tilde{A}_i=\begin{pmatrix}
				\tilde{A}_{q,i}&\tilde{B}_{q,i}\Gamma_{i}\\0&S_{i}
			\end{pmatrix}, \tilde{B}_i=\begin{pmatrix}
				\tilde{B}_{q,i}D_{p,i}\\B_{p,i}
			\end{pmatrix},\tilde{E}_i=\begin{pmatrix}
				\tilde{E}_{q,i}\\0
			\end{pmatrix}, \tilde{C}_i=\begin{pmatrix}
				\tilde{C}_{q,i}&0
			\end{pmatrix}. 
		\end{equation*}
		It is proved that Assumption A3.2 implies $(\tilde{C}_i, \tilde{A}_i)$ is detectable in Appendix \ref{app11}. 
		
		Then, the cascade system of pre-compensator \eqref{compensator1a2} and agent system \eqref{hete_sys} can be written as follows,
		\begin{equation}\label{hekf-system-ana2}
			\begin{system}{cl}
				{\tilde{x}}_i(k+1) &=\tilde{A}_i\tilde{x}_i(k)+\tilde{B}_iv_i(k)+\tilde{E}_i\omega_i(k),\\
				y_i(k)&=\tilde{C}_i\tilde{x}_i(k)+{G}_i\omega_i(k),\\
				{z}_i(k)&=\begin{pmatrix}
					\tilde{C}_{q,i}^m&0
				\end{pmatrix}\tilde{x}_i(k), 
			\end{system}
		\end{equation}
		where $\tilde{x}_{i}(k)\in\R^{\tilde{n}_{i}}$, $v_{i}(k)\in\R^{p}$, $y_{i}(k)\in\R^{p}$ are states, inputs and outputs of the cascade system. We have ($\tilde{A}_i,\tilde{B}_i$) stabilizable, which is proved in Appendix \ref{app01}.

		{We see that,
			$(\tilde{A}_i, \tilde{C}_i, \tilde{E}_i, \tilde{G}_i)$ satisfies
			\begin{align*}
				&\tilde{\Pi}_i
				{S}_i=\tilde{A}_i\tilde{\Pi}_i+\tilde{E}_i,\\
				&\tilde{C}_i\tilde{\Pi}_i+
				{G}_i=0,
			\end{align*}
			with
			$
			\tilde{\Pi}_i=\begin{pmatrix}
				\Pi_{i}\\I_{n_{\omega_i}}
			\end{pmatrix}$. Meanwhile, $(\tilde{A}_i, \tilde{B}_i, \tilde{C}_i)$ is of uniform rank $\rho_i\geq 1$.}

		Now by defining $\bar{x}_i(k)=\tilde{x}_i(k)-\tilde{\Pi}_i \omega_i(k)$, we can obtain
		\begin{align*}
			{\bar{x}}_i(k+1)=&{\tilde{x}}_i(k+1)-\tilde{\Pi}_i{\omega}_i(k+1)\\
			=&\tilde{A}_i\tilde{x}_i(k)+\tilde{B}_iv_i(k)+\tilde{E}_i\omega_i(k)-\tilde{\Pi}_i{S}_i\omega_i(k)\\
			=&\tilde{A}_i\tilde{x}_i(k)+\tilde{B}_iv_i(k)+\tilde{E}_i\omega_i(k)-\left(\tilde{A}_i\tilde{\Pi}_i+\tilde{E}_i\right)\omega_i(k)\\
			=&\tilde{A}_i\bar{x}_i(k)+\tilde{B}_iv_i(k),\\
			{y}_i(k)=&\tilde{C}_i\tilde{x}_i(k)-\tilde{C}_i\tilde{\Pi}_i \omega_i(k)+\tilde{C}_i\tilde{\Pi}_i \omega_i(k)+{G}_i\omega_i(k)\\
			=&\tilde{C}_i\bar{x}_i(k)+(\tilde{C}_i\tilde{\Pi}_i+{G}_i)\omega_i(k)\\
			=&\tilde{C}_i\bar{x}_i(k).
		\end{align*}
		Let  $\tilde{C}_i^m=\begin{pmatrix}
			\tilde{C}_{q,i}^m&-\tilde{C}_{q,i}^m\Pi_{i}
		\end{pmatrix}$, which implies $\tilde{C}_i^m\tilde{\Pi}_i=0$. Then, we can define a new measurement
		\begin{align*}
			\bar{z}_i(k)&=z_i(k)-\tilde{C}_{q,i}^m\Pi_{i}p_i(k)\\
			&=\tilde{C}_{q,i}^m \eta_i(k)-\tilde{C}_{q,i}^m\Pi_{i}p_i(k)\\
			&=\tilde{C}_i^m\tilde{x}_i(k)-\tilde{C}_i^m\tilde{\Pi}_i\omega_i(k)\\
			&=\tilde{C}_i^m\bar{x}_i(k).
		\end{align*}
		Then, the above system can be rewritten as,
		\begin{equation}\label{hekf-systema2}
			\begin{system}{cl}
				{\bar{x}}_i(k+1) &=\tilde{A}_i\bar{x}_i(k)+\tilde{B}_iv_i(k),\\
				{y}_i(k)&=\tilde{C}_i\bar{x}_i(k),\\
				\bar{z}_i(k)&=\tilde{C}_i^m\bar{x}_i(k),
			\end{system}
		\end{equation}
		where $\bar{x}_{i}(k)\in\R^{n_{i}}$, $v_{i}(k)\in\R^{p}$, $\bar{z}_{i}(k)\in\R^{z_i}$ are states, inputs and measurement outputs. And, ($\tilde{C}_i, \tilde{A}_i$) and ($\tilde{C}_i^m, \tilde{A}_i$) are detectable according to the proof in Appendix \ref{app2}.

		\begin{remark}
			In the compensated system \eqref{hekf-systema2}, the new measurement $\bar{z}_i(k)$ consists of original $z_i(k)$ and the pre-compensator II's state $p_i(k)$. Because 
			\[
			\tilde{C}_i^m\tilde{x}_i(k)=\begin{pmatrix}
				\tilde{C}_{q,i}^m&-\tilde{C}_{q,i}^m\Pi_{i}
			\end{pmatrix}\begin{pmatrix}
				\eta_i(k)\\p_i(k)
			\end{pmatrix}=z_i(k)-\tilde{C}_{q,i}^m\Pi_{i}p_i(k),
			\]
			it can be achieved via using $z_i(k)$ and measuring $p_i(k)$. 
			
			Moreover,  ($\tilde{C}_{q,i}^m, \tilde{A}_{q,i}$) is detectable under Assumption A3.4, which is proved in Appendices \ref{app11} and \ref{app2}.
		\end{remark}
		\vspace{1mm}

			\noindent\textbf{Step 3: Remodeling the reference exosystem }
			
			Following the design procedure in \cite{wang-saberi-yang}, we can remodel the reference exosystem \eqref{exo} into the target system $(C_h,A_h,B_h)$, which is given in the following Lemma~\ref{lem-exo}. This lemma is from \cite[Appendix A.2]{wang-saberi-yang} and will be used to shape the reference exosystem.
			\begin{lemma}\label{lem-exo}
				For the given exosystem \eqref{exo}, there exists another target system that is defined as follows,
				\begin{equation}\label{exo-2}
					\begin{system}{cl}
						{x}_h(k+1)&=A_h{x}_h(k), \quad {x}_{h}(k)\in \mathbb{R}^{{h}}, {x}_h(0)={x}_{h0},\\
						y_r(k)&=C_h{x}_h(k),
					\end{system}
				\end{equation}
				such that for all $x_{r0} \in \mathbb{R}^r$, there exists ${x}_{h0}\in \mathbb{R}^{{h}}$ for which \eqref{exo-2} {creates} exactly the same output $y_r(k)$ as the original exosystem \eqref{exo}. Moreover, a matrix $B_h$ can be found such that the triple $(C_h,A_h,B_h)$ is invertible, of uniform rank $n_q$, and has no invariant zeros, where $n_q$ is an integer greater than or equal to maximal order of infinite zeros of $(C_i,A_i,B_i),\  i\in \{1,\cdots,N\}$ and all the observability indices (see \cite{chen-lin-shamash} for the definition) of $(C_r, A_r)$. Note that the eigenvalues of $A_h$ include all eigenvalues of $A_r$ and additional zero eigenvalues. 
			\end{lemma}		
			\vspace{1mm}


			\noindent\textbf{Step 4: Designing Pre-compensator III to almost homogenize the compensated MAS} 
			
			According to \cite[Appendix A.1]{wang-saberi-yang}, a pre-compensator can be developed for each system \eqref{hekf-systema2} ($i=1,\ldots,N$) to almost homogenize it to the target model $(C_h,A_h,B_h)$. The pre-compensator is of the form,
			\begin{equation}\label{pre_con}
				\begin{system}{cl}
					{\xi}_i(k+1)&=A_{i,h}\xi_i(k)+B_{i,h}\bar{z}_i(k)+E_{i,h}\check{u}_i(k),\\
					v_i(k)&=C_{i,h}\xi_i(k)+D_{i,h}\check{u}_i(k)+F_{i,h}\bar{z}_i(k),
				\end{system}
			\end{equation} 
			such that the cascade system of this pre-compensator and system \eqref{hekf-systema2} can be written as
			\begin{equation}\label{sys_homo}
				\begin{system}{cl}
					{\check{x}}_i(k+1)&=A_h\check{x}_i(k)+B_h(\check{u}_i(k)+\phi_i(k)),\\
					{y}_i(k)&=C_h\check{x}_i(k),
				\end{system}
			\end{equation} 
			where $\phi_i(k)$ is generated by 
			\begin{equation}\label{sys-rho}
				\begin{system}{cl}
					{\sigma}_i(k+1)&=A_{i,s}\sigma_i(k),\\
					\phi_i(k)&=C_{i,s}\sigma_i(k),
				\end{system}
			\end{equation}
			and $A_{i,s}$ is Schur stable. \\

			\noindent\textbf{Step 5: Designing a scalable protocol for compensated agent $\check{x}_i(k)$}
			
			Select matrices $H$ and $K$ to have $A_h-HC_h$ and $A_h-B_hK$ Schur stable. Then, with the localized information exchange,  a scalable protocol is developed for each compensated agent \eqref{sys_homo}, i.e.,
			\begin{equation}\label{pscp2}
				\begin{system}{cl}
					{\hat{x}}_i(k+1)&=A_h\hat{x}_i(k)-B_hK\hat{\zeta}_i(k)+H(\tilde{\zeta}_i(k)-C_h\hat{x}_i(k))-\frac{\iota_i}{2+\bar{d}_{\text{in}}(i)}B_hK\chi_i(k)\\
					{\chi}_i(k+1)&=A_h\chi_i(k)+B_h\check{u}_i(k)+\hat{x}_i(k)-\hat{\zeta}_i(k)-\frac{\iota_i}{2+\bar{d}_{\text{in}}(i)}\chi_i(k)\\
					\check{u}_i(k)&=-K\chi_i(k)
				\end{system}
			\end{equation}
			where 
			\begin{equation}\label{hatzetahete2}
				\hat{\zeta}_i(k)=\sum_{j=1}^{N}a_{ij}(\chi_i(k)-\chi_j(k)),
			\end{equation}
			with $\chi_i(k)$ being the internal states in \eqref{etahat}, i.e., $\eta_i(k)=\chi_i(k)$.\\
			
			\noindent\textbf{Step 6: Obtaining an overall scalable protocol for agent $i$}
			
			By combining pre-compensator \eqref{precomz} on Step $1$, pre-compensator \eqref{compensator1a2} on Step $2$, pre-compensator \eqref{pre_con} on Step~$4$, and scalable protocol \eqref{pscp2} on Step $5$, the overall scalable protocol for agent $i$ is obtained,
			\begin{equation}\label{pscp2final2}
				\begin{system}{cl}
					{q}_i(k+1)&=A_{q,i}q_i(k)+B_{q,i}(\Gamma_{i}-D_{p,i}F_{i,h}\tilde{C}_{q,i}^m\Pi_i)p_i(k)+B_{q,i}D_{p,i}C_{i,h}\xi_i(k)\\
					&\hspace{3cm}-B_{q,i}D_{p,i}D_{i,h}K\chi_i(k)+B_{q,i}D_{p,i}F_{i,h}z_i(k)\\
					{p}_i(k+1)&=(S_i-B_{p,i}F_{i,h}\tilde{C}_{q,i}^m\Pi_i)p_i(k)+B_{p,i}C_{i,h}\xi_i(k)\\
					&\hspace{3cm}-B_{p,i}D_{i,h}K\chi_i(k)+B_{p,i}F_{i,h}z_i(k)\\
					{\xi}_i(k+1)&=A_{i,h}\xi_i+B_{i,h} z_i(k)-B_{i,h} \tilde{C}_{q,i}^m\Pi_ip_i(k)-E_{i,h}K\chi_i(k),\\
					{\hat{x}}_i(k+1)&=A_h\hat{x}_i(k)-B_hK\hat{\zeta}_i(k)+H(\tilde{\zeta}_i(k)-C_h\hat{x}_i(k))-\frac{\iota_i}{2+\bar{d}_{\text{in}}(i)}B_hK\chi_i(k)\\
					{\chi}_i(k+1)&=A_h\chi_i(k)-B_hK\chi_i(k)+\hat{x}_i(k)-\hat{\zeta}_i(k)-\frac{\iota_i}{2+\bar{d}_{\text{in}}(i)}\chi_i(k)\\
					u_i(k)&=C_{q,i}q_i(k)+D_{q,i}(\Gamma_{i}-D_{p,i}F_{i,h}\tilde{C}_{q,i}^m\Pi_i)p_i(k)+D_{q,i}D_{p,i}C_{i,h}\xi_i(k)\\
					&\hspace{3cm}-D_{q,i}D_{p,i}D_{i,h}K\chi_i(k)+D_{q,i}D_{p,i}F_{i,h}z_i(k).
				\end{system}
			\end{equation}

		Then, we state the main result for scalable regulated output synchronization of heterogeneous MAS in the presence of disturbances with known frequencies as follows.
		\begin{theorem}\label{thm_f1heter}
			Consider a heterogeneous MAS modeled by \eqref{hete_sys}, \eqref{local}, and \eqref{zetabar} in the presence of disturbances generated by \eqref{hekf-distsys2}, and a reference trajectory generated by \eqref{exo}. Let Assumptions \ref{ass-exo} and \ref{ass1} hold. 
			
			Then,
			the scalable regulated output synchronization problem defined in Problem \ref{prob_reg_xheter} is solvable. In particular, the protocol \eqref{pscp2final2} achieves  regulated output synchronization \eqref{reg_synch_state} among agents for any $N$, any fixed graph
			$\mathscr{G}\in\mathbb{G}^N_\mathscr{C}$, any process disturbances and measurement noise generated by \eqref{hekf-distsys2}, and any reference trajectory given by \eqref{exo}.
		\end{theorem} 
		
		\begin{proof}[The proof of Theorem \ref{thm_f1heter}]
			The closed-loop expression of cascade system \eqref{sys_homo}, scalable protocol \eqref{pscp2final2}, relative information \eqref{zetabar2}, and local information exchange \eqref{hatzetahete2} are rewritten as
			\begin{equation}\label{newsystem1}
				\begin{system}{cl}
					{\check{x}}_i(k+1) &={A}_h\check{x}_i(k)-{B}_hK\chi_i(k)+{B}_hC_{i,s}\sigma_i(k),\\
					{\hat{x}}_i(k+1)&=A_h\hat{x}_i(k)+HC_h(\frac{1}{2+\bar{d}_{\text{in}}(i)}\sum_{j=1}^N\tilde{\ell}_{ij}\check{x}_j(k)-\hat{x}_i(k))\\
					&\hspace{3cm}-\frac{1}{2+\bar{d}_{\text{in}}(i)}B_hK\sum_{j=1}^N\tilde{\ell}_{ij}\chi_j(k),\\
					{\chi}_i(k+1)&=A_h\chi_i(k)-B_hK\chi_i(k)+\hat{x}_i(k)-\frac{1}{2+\bar{d}_{\text{in}}(i)}\sum_{j=1}^N\tilde{\ell}_{ij}\chi_j(k),\\
					{\sigma}_i(k+1)&=A_{i,s}\sigma_i(k),\\
					{y}_i(k)&={C}_h\check{x}_i(k),
				\end{system}
			\end{equation}
			for $i=1,\ldots,N$.  Then, by defining $\bar{\tilde{x}}_i(k)=\check{x}_i(k)-x_h(k)$, $\bar{\tilde{y}}_i(k)=y_i(k)-y_r(k)$, we obtain that
			\begin{equation}\label{newsystem4}
				\begin{system}{cl}
					{\bar{\tilde{x}}}_i(k+1) &={A}_h\bar{\tilde{x}}_i(k)-{B}_hK\chi_i(k)+{B}_hC_{i,s}\sigma_i(k),\\
					{\hat{x}}_i(k+1)&=A_h\hat{x}_i(k)+HC_h(\frac{1}{2+\bar{d}_{\text{in}}(i)}\sum_{j=1}^N\tilde{\ell}_{ij}\bar{\tilde{x}}_j(k)-\hat{x}_i(k))\\
					&\hspace{3cm}-\frac{1}{2+\bar{d}_{\text{in}}(i)}B_hK\sum_{j=1}^N\tilde{\ell}_{ij}\chi_j(k),\\
					{\chi}_i(k+1)&=A_h\chi_i(k)-B_hK\chi_i(k)+\hat{x}_i(k)-\frac{1}{2+\bar{d}_{\text{in}}(i)}\sum_{j=1}^N\tilde{\ell}_{ij}\chi_j(k),\\
					{\sigma}_i(k+1)&=A_{i,s}\sigma_i(k),\\
					\bar{\tilde{y}}_i(k)&={C}_h\bar{\tilde{x}}_i(k),
				\end{system}
			\end{equation}
			for $i=1,\ldots,N$. 
			Similarly, we only just need to prove the above system \eqref{newsystem4} is stable. 
			
			Now, let
			\begin{equation*}
				\bar{\tilde{x}}(k)=\begin{pmatrix}
					\bar{\tilde{x}}_1(k)\\ \vdots\\ \bar{\tilde{x}}_N(k)
				\end{pmatrix},
				\hat{x}(k)=\begin{pmatrix}
					\hat{x}_1(k)\\ \vdots\\ \hat{x}_N(k)
				\end{pmatrix},
				{\chi}(k)=\begin{pmatrix}
					{\chi}_1(k)\\ \vdots\\ {\chi}_N(k)
				\end{pmatrix},\sigma(k)=\begin{pmatrix}
					\sigma_1(k)\\ \vdots\\ \sigma_N(k)
				\end{pmatrix},\bar{\tilde{y}}(k)=\begin{pmatrix}
					\bar{\tilde{y}}_1(k)\\ \vdots\\\bar{\tilde{y}}_N(k)\end{pmatrix}.
			\end{equation*}
			Then, we obtain the following system
			\begin{equation}
				\begin{system}{rl}
					{\bar{\tilde{x}}}(k+1) &=(I\otimes{A}_h)\bar{\tilde{x}}(k)-(I\otimes{B}_hK)\chi(k)+(I\otimes{B}_h)C_s\sigma(k),\\
					{\hat{x}}(k+1)&=(I\otimes({A}_h-HC_h))\hat{x}(k)-((I-\tilde{D})\otimes{B}_hK)\chi(k)+((I-\tilde{D})\otimes HC_h)\bar{\tilde{x}}(k)\\
					{\chi}(k+1)&=(\tilde{D}\otimes {A}_h-I\otimes(B_hK))\chi(k)+(I\otimes{A}_h)\hat{x}(k)\\
					{\sigma}(k+1)&=A_{s}\sigma(k)\\
					\bar{\tilde{y}}(k)&=(I\otimes{C}_h)\bar{\tilde{x}}(k),
				\end{system}
			\end{equation}
			with $A_{s}=\diag(A_{i,s})$ and $C_s=\diag(C_{i,s})$.

			Define errors $e(k)=\bar{\tilde{x}}(k)-{\chi}(k)$ and $\bar{e}(k)=((I-\tilde{D})\otimes I)\bar{\tilde{x}}(k)-\hat{x}(k)$. We have the overall error system
			\begin{equation}\label{newsystem3}
				\begin{system}{cl}
					{\bar{\tilde{x}}}(k+1) &=(I\otimes({A}_h-{B}_hK))\bar{\tilde{x}}(k)+(I\otimes{B}_hK)e(k)+(I\otimes{B}_h)C_s\sigma(k),\\
					{\bar{e}}(k+1)&=(I\otimes ({A}_{h}-HC_h))\bar{e}(k)+((I-\tilde{D})\otimes{B}_h)C_s\sigma(k),\\
					{e}(k+1)&=(\tilde{D}\otimes {A}_h)e(k)+\bar{e}(k)+(I\otimes{B}_h)C_s\sigma(k),\\
					{\sigma}(k+1)&=A_{s}\sigma(k),\\
					\bar{\tilde{y}}(k)&=(I\otimes{C}_h)\bar{\tilde{x}}(k).
				\end{system}
			\end{equation}
			
			Because all eigenvalues of $A_h$ are on the unit circle and all eigenvalues of $\tilde{D}$ are in the open unit circle,  all eigenvalues of $\tilde{D}\otimes {A}_h$ are in the open unit circle.  Moreover, $A_s$, ${A}_h-{B}_hK$, and ${A}_{h}-HC_h$ are Schur stable. Hence, $\bar{\tilde{x}}(k)$ is asymptotically stable, i.e.  
			\[
			\lim_{k\to\infty}\bar{\tilde{x}}_i(k)\to 0,
			\]
			which implies that $\bar{\tilde{y}}_i(k)=C_h\bar{\tilde{x}}_i(k)\to 0$, i.e. $y_i(k)\to {y}_r(k)$ as $k\to \infty$.			
		\end{proof}
		
		\section{Numerical examples}
		
		In this section, two examples are created to illustrate the effectiveness and scalability of our scale-free protocol design for homogeneous and heterogeneous MAS with known frequency disturbances.

		\subsection{Homogeneous MAS}
		The agent system in the form of \eqref{homo_sys} is chosen as
		\begin{align*}
			A=\begin{pmatrix}
				1&1&0\\
				0&1&1\\
				0&0&1
			\end{pmatrix},\ B=\begin{pmatrix}
				0\\
				0\\
				1
			\end{pmatrix},\ E=\begin{pmatrix}
				1&0\\
				0&1\\
				1&1
			\end{pmatrix},\\
			C=\begin{pmatrix}
				1&0&0
			\end{pmatrix},\ 	G=\begin{pmatrix}
				1&1
			\end{pmatrix}.
		\end{align*}
		with the disturbance model by
		\[
		S=\begin{pmatrix}
			0&1\\
			-1&0
		\end{pmatrix}.
		\]
		Solving equations \eqref{regeq}, we obtain matrices $\Pi$ and $\Gamma$ as
		\[
		\Pi=\begin{pmatrix}
			-1&-1\\
			1&0\\
			0&0
		\end{pmatrix}, \qquad \Gamma=\begin{pmatrix}
			-1&-1
		\end{pmatrix},
		\]
		which gives pre-compensator \eqref{compensator1a} with
		$
		B_p=\begin{pmatrix}
			0\\1
		\end{pmatrix}$,  $D_p=0$.

		Meanwhile, choose $K=\begin{pmatrix}
			-0.1& -1 &-2& 0.1& 2
		\end{pmatrix}$,	$H=\begin{pmatrix}
			2&1&0.5&-0.5&0
		\end{pmatrix}\T$ 
		such that $\tilde{A}-\tilde{B}K$ and $\tilde{A}-H\tilde{C}$ are Schur stable. So, we can obtain the following protocol		
		\begin{equation}\label{pscp2sim}
			\begin{system}{cl}
				{p}_i(k+1)&=\begin{pmatrix}
					0&1\\-1&0
				\end{pmatrix}p_i(k)-\begin{pmatrix}
					0&0&0&0&0\\
					-0.1&   -1&   -2&    0.1&    2
				\end{pmatrix}\chi_i(k),\\
				{\hat{x}}_i(k+1)&= \begin{pmatrix}
					-1  &   1   &  0  &   0  &   0\\
					-1  &   1   &  1  &   0  &   0\\
					-0.5  &   0   &  1  &   -1  &   -1\\
					0.5  &   0   &  0  &   0  &   1\\
					0  &   0   &  0  &  -1  &   0
				\end{pmatrix}   \hat{x}_i(k)+\begin{pmatrix}
					2\\1\\0.5\\-0.5\\0
				\end{pmatrix}{\zeta}_i(k)-\begin{pmatrix}
					0  &   0   &  0  &   0  &   0\\
					0  &   0   &  0   &  0  &   0\\
					0  &   0  &   0  &  0  &   0\\
					0  &   0  &   0  &  0  &   0\\
					-0.1&   -1&   -2&    0.1&    2
				\end{pmatrix}
				\hat{\zeta}_i(k)\\
				{\chi}_i(k+1)&=\begin{pmatrix}
					1  &   1   &  0  &   0  &   0\\
					0  &   1  &   1  &   0   &  0\\
					0  &   0  &   1  &   -1   &  -1\\
					0  &   0  &  0  & 0  &  1\\
					0.1  &   1  &  2  & -1.1  &  2
				\end{pmatrix}\chi_i(k)+\begin{pmatrix}
					1  &   1   &  0  &   0  &   0\\
					0  &   1  &   1  &   0   &  0\\
					0  &   0  &   1  &   -1   &  -1\\
					0  &   0  &  0  & 0  &  1\\
					0  &   0  &  0  & -1  &  0
				\end{pmatrix}(\hat{x}_i(k)-\hat{\zeta}_i(k))\\
				u_i(k)&=-\begin{pmatrix}
					1&1
				\end{pmatrix} p_i(k)
			\end{system}
		\end{equation}
		
		Next, we will show that the same scalable protocol \eqref{pscp2sim} can achieve output synchronization for both 6 and 60 agents with different communication topologies.
		
		\subsection*{\bf Case I: 6-agent graph in homogeneous MAS}
		A homogeneous MAS with $6$ agents has a directed communication topology with its associate adjacency matrix $\mathcal{A}_{I}$ being with $a_{21}=a_{32}=a_{13}=a_{43}=a_{36}=a_{54}=a_{65}=1$. 
		
		
		\begin{figure}[h!]
			\includegraphics[width=13cm]{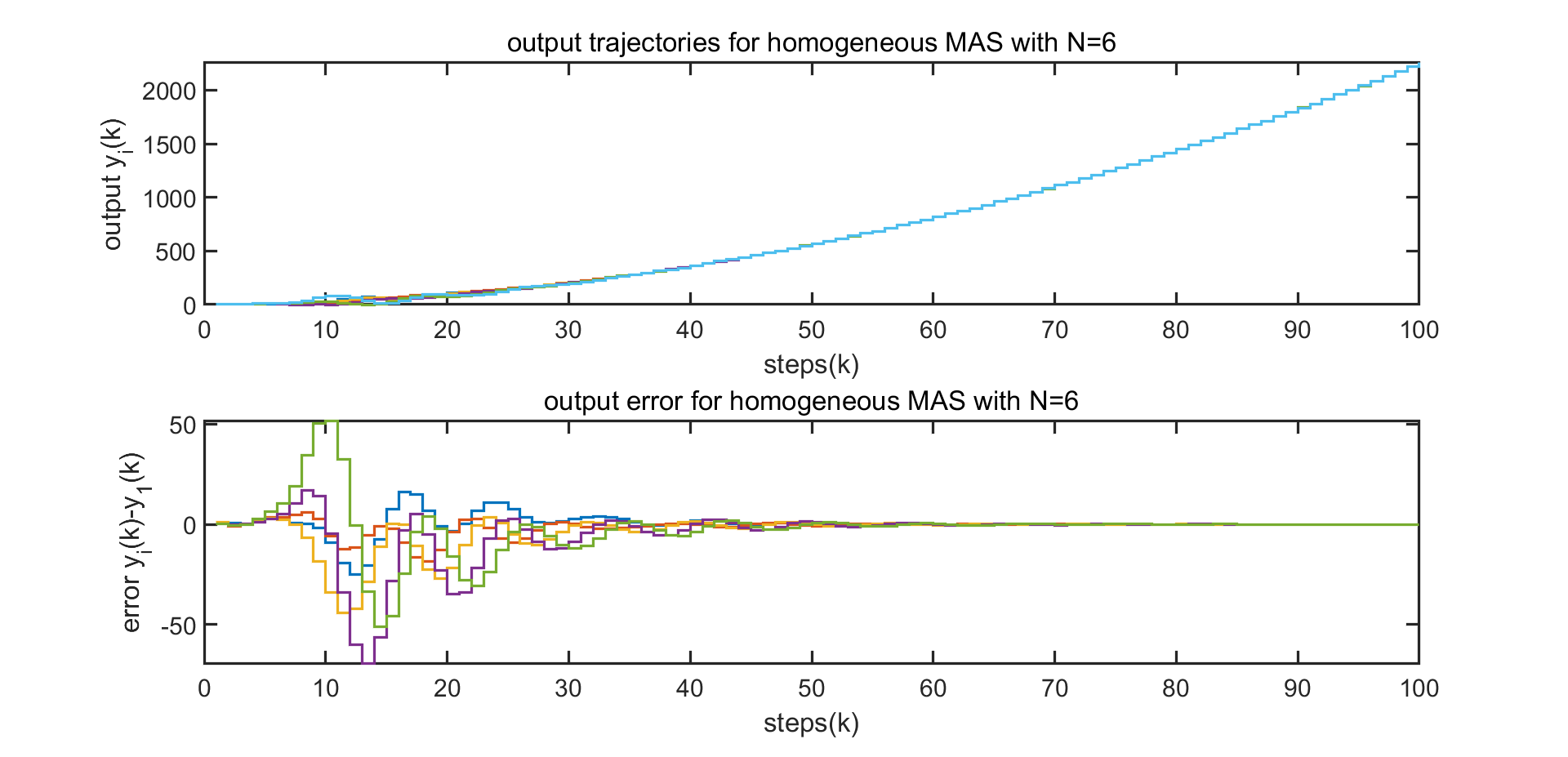}\centering
			\caption{{Scalable exact output synchronization for homogeneous discrete-time 
					MAS with communication topology described in Case I.}}\label{results_case2.1}
		\end{figure}
		
		\subsection*{\bf Case II: 60-agent graph in homogeneous MAS}
		A homogeneous MAS with $60$ agents has a directed loop communication topology with its associate adjacency matrix $\mathcal{A}_{II}$ being with $a_{i+1,i}=a_{1,60}=1$ and $i=1,\cdots,59$. 
		
		
		\begin{figure}[h!]
			\includegraphics[width=13cm]{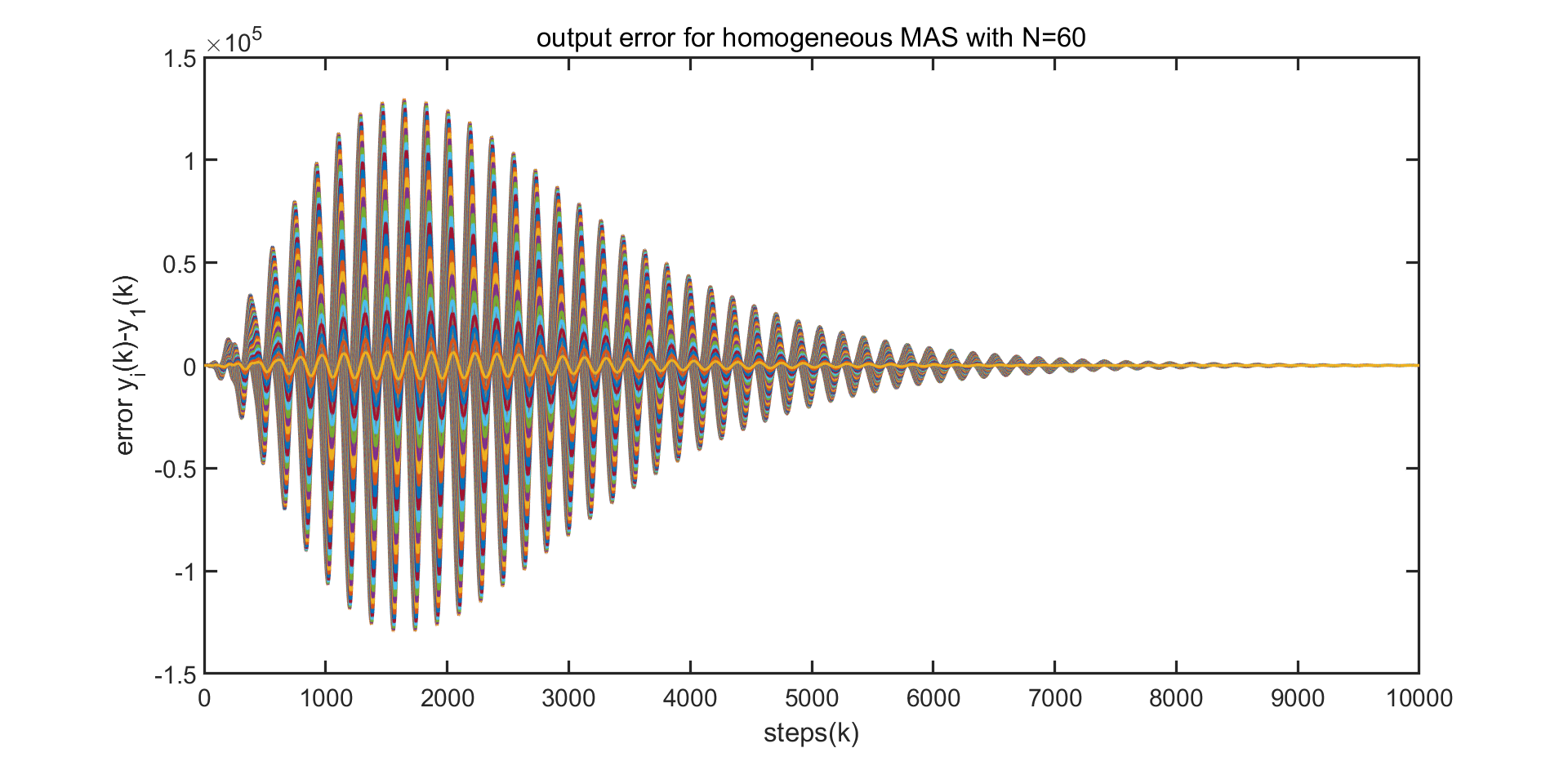}\centering
			\caption{{Scalable exact output synchronization for homogeneous discrete-time 
					MAS with communication topology described in Case II.}}\label{results_case3.1}
		\end{figure}
		
		It is shown in Figure \ref{results_case2.1} and Figure \ref{results_case3.1} that scalable output synchronization is achieved for a homogeneous discrete-time  MAS with  both $N=6$ agents and $N=60$ agents. 
		\subsection{Heterogeneous MAS}
		The non-identical agent \eqref{hete_sys} with disturbance \eqref{hekf-distsys2} satisfying Assumption \ref{ass1} can be chosen from the following three models:
		\begin{equation*}
			\text{Group 1:}\begin{system}{l}
				A_i=\begin{pmatrix}
					0&0\\0&-1
				\end{pmatrix}, B_i=I_2, E_i=I_2, C_i\T=\begin{pmatrix}
					1\\1
				\end{pmatrix},\\
				G_i=\begin{pmatrix}
					1&0
				\end{pmatrix}, C^m_i=I_2, S_i=\begin{pmatrix}
					0&1\\-1&0
				\end{pmatrix},
			\end{system}
		\end{equation*}
		\begin{equation*}
			\text{Group 2:}\begin{system}{l}
				A_i=\begin{pmatrix}
					0&1\\-1&0
				\end{pmatrix},B_i=I_2, E_i=\begin{pmatrix}
					1\\0
				\end{pmatrix}, C_i\T=\begin{pmatrix}
					1\\1
				\end{pmatrix},\\
				G_i=2, C^m_i=I_2, S_i=1,
			\end{system}
		\end{equation*}
		and
		\begin{equation*}
			\text{Group 3:}\begin{system}{l}
				A_i=\begin{pmatrix}
					0&1\\-3&-4
				\end{pmatrix}, B_i=\begin{pmatrix}
					0\\1
				\end{pmatrix}, E_i=I_2, C_i\T=\begin{pmatrix}
					1\\2
				\end{pmatrix},\\
				G_i=\begin{pmatrix}
					1&0
				\end{pmatrix}, C^m_i=I_2, S_i=\begin{pmatrix}
					0&2\\-2&0
				\end{pmatrix}.
			\end{system}
		\end{equation*}
		Moreover,  the exosystem \eqref{exo} is selected as,
		\[
		A_r=\begin{pmatrix}
			1&1&0\\0&1&1\\0&0&1
		\end{pmatrix}, C_r=\begin{pmatrix}
			1&0&0
		\end{pmatrix}.
		\]
		Since $n_q=3$, we only remodel target system by $A_h=A_r$, $B_h=\begin{pmatrix}
			0&0&1
		\end{pmatrix}\T$, and $C_h=C_r$.

		Next, we design the following Pre-compensator I, II, and III for the above Agent Group $1$, $2$, and $3$, respectively.
		\[\text{Group 1:}
		\begin{system}{l}
			u_i(k)=\begin{pmatrix}
				1\\-1
			\end{pmatrix}u_i^q(k),\\
			{p}_i(k+1)=\begin{pmatrix}
				0&1\\-1&0
			\end{pmatrix}p_i(k)+\begin{pmatrix}
				2\\1
			\end{pmatrix}v_i(k)\\
			{u}_i^q(k)=\begin{pmatrix}
				1&-2
			\end{pmatrix}p_i(k)-5v_i(k),\\
			{\xi}_i(k+1)=\xi_i(k)+5\check{u}_i(k),\\
			v_i(k)=\xi_i(k)-\begin{pmatrix}
				0&1
			\end{pmatrix}\bar{z}_i(k),
		\end{system}
		\]
		with $\Pi_i=\begin{pmatrix}
			-2&-2\\1&2
		\end{pmatrix}$.
		\[\text{Group 2:}
		\begin{system}{l}
			u_i(k)=\begin{pmatrix}
				2\\1
			\end{pmatrix}u_i^q(k),\\
			{p}_i(k+1)=p_i(k)-v_i(k)\\
			{u}_i^q(k)=-2p_i(k)+2v_i(k)\\
			{\xi}_i(k+1)=\begin{pmatrix}
				1 &   1 \\
				0 &   1 
			\end{pmatrix}\xi_i(k)
			+\begin{pmatrix}
				0\\
				6
			\end{pmatrix}\check{u}_i(k),\\
			v_i(k)=\begin{pmatrix}
				1   &  0
			\end{pmatrix}\xi_i(k)-
			\begin{pmatrix}
				0  &  \frac{2}{3}
			\end{pmatrix}
			\bar{z}_i(k),
		\end{system}
		\]
		with $\Pi_i=\begin{pmatrix}
			-2.5&0.5
		\end{pmatrix}\T$.
		\[\text{Group 3:}
		\begin{system}{l}
			{p}_i(k+1)=\begin{pmatrix}
				0&2\\-2&0
			\end{pmatrix}p_i(k)+\begin{pmatrix}
				\frac{1}{37}\\-\frac{1}{39}
			\end{pmatrix}v_i(k)\\
			{u}_i(k)=-\begin{pmatrix}
				\frac{37}{17}&\frac{39}{17}
			\end{pmatrix}p_i(k)-\frac{76}{1443}v_i(k)\\
			{\xi}_i(k+1)=\begin{pmatrix}
				1 &   1 \\
				0 &   1 
			\end{pmatrix}\xi_i(k)
			-\begin{pmatrix}
				0\\
				0.1053
			\end{pmatrix}\check{u}_i(k),\\
			v_i(k)=\begin{pmatrix}
				1   &  0
			\end{pmatrix}\xi_i(k)+
			\begin{pmatrix}
				17  &  -\frac{17}{4}
			\end{pmatrix}
			\bar{z}_i(k),
		\end{system}
		\]
		with $\Pi_i=\frac{1}{17}\begin{pmatrix}
			1&-4\\-9&2
		\end{pmatrix}$.
		
		Then, we choose $K=\begin{pmatrix}
			0.1&  1  &  2
		\end{pmatrix}$ and $H=\begin{pmatrix}
			2  &  1 & 0.5
		\end{pmatrix}$ such that $A_h-B_h K$ and $A_h-HC_h$ are Schur stable.
		
		
		Next, we show that the above protocols can achieve regulated output synchronization for both 6 and 60 agents with different communication topologies. Meanwhile, let $\iota_1=1$ and $\iota_i=0$ for $i\neq 1$.

		\subsection*{\bf Case I: 6-agent graph in heterogeneous MAS}
		
		Consider a heterogeneous MAS consisting of $6$ agents with $(C_i, A_i, B_i)$ for $i=1,6 \in\text{Group 1}$, $i=2,5 \in\text{Group 2}$, and $i=3,5 \in\text{Group 3}$. The associated adjacency matrix is denoted by $\mathcal{A}_{I}$ with $a_{21}=a_{32}=a_{13}=a_{43}=a_{36}=a_{54}=a_{65}=1$. 
		
		\subsection*{\bf Case II: 60-agent graph in heterogeneous MAS} 
		
		consider a heterogeneous MAS consisting of $60$ agents with $(C_i, A_i, B_i)$ for $i=1,\ldots,20 \in\text{Group 1}$, $i=21,\ldots,40 \in\text{Group 2}$, and $i=41,\ldots,60 \in\text{Group 3}$, and the communication topology with associated adjacency matrix $\mathcal{A}_{II}$ being $a_{i+1,i}=a_{1,60}=1$ and $i=1,\cdots,59$. 
		
		\begin{figure}[h!]
			\includegraphics[width=13cm]{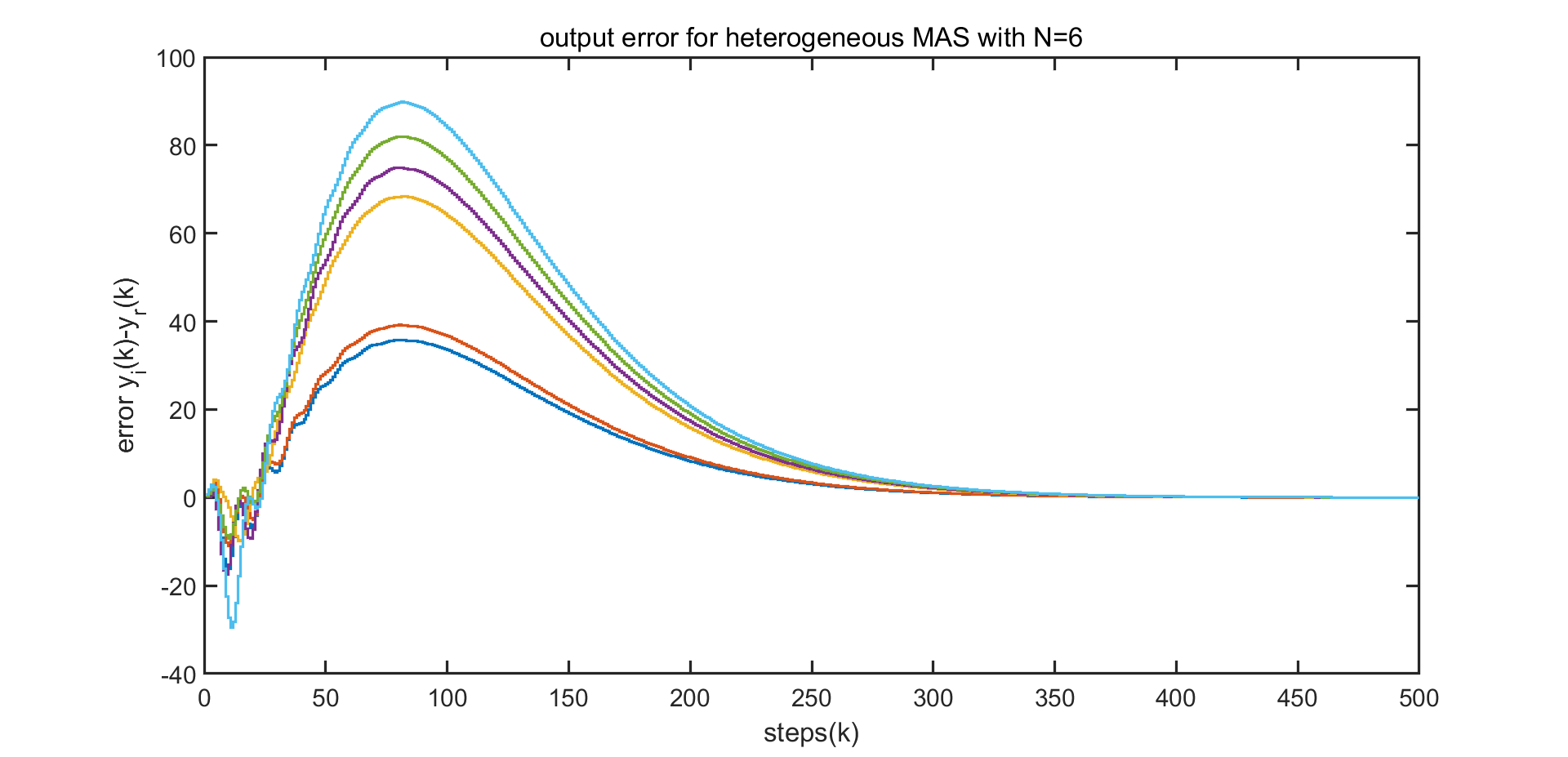}
			\centering
			\caption{Scalable exact regulated output synchronization for heterogeneous
				discrete-time MAS with communication topology described in Case I.}\label{het_N6}
		\end{figure}
		\begin{figure}[h!]
			\includegraphics[width=13cm]{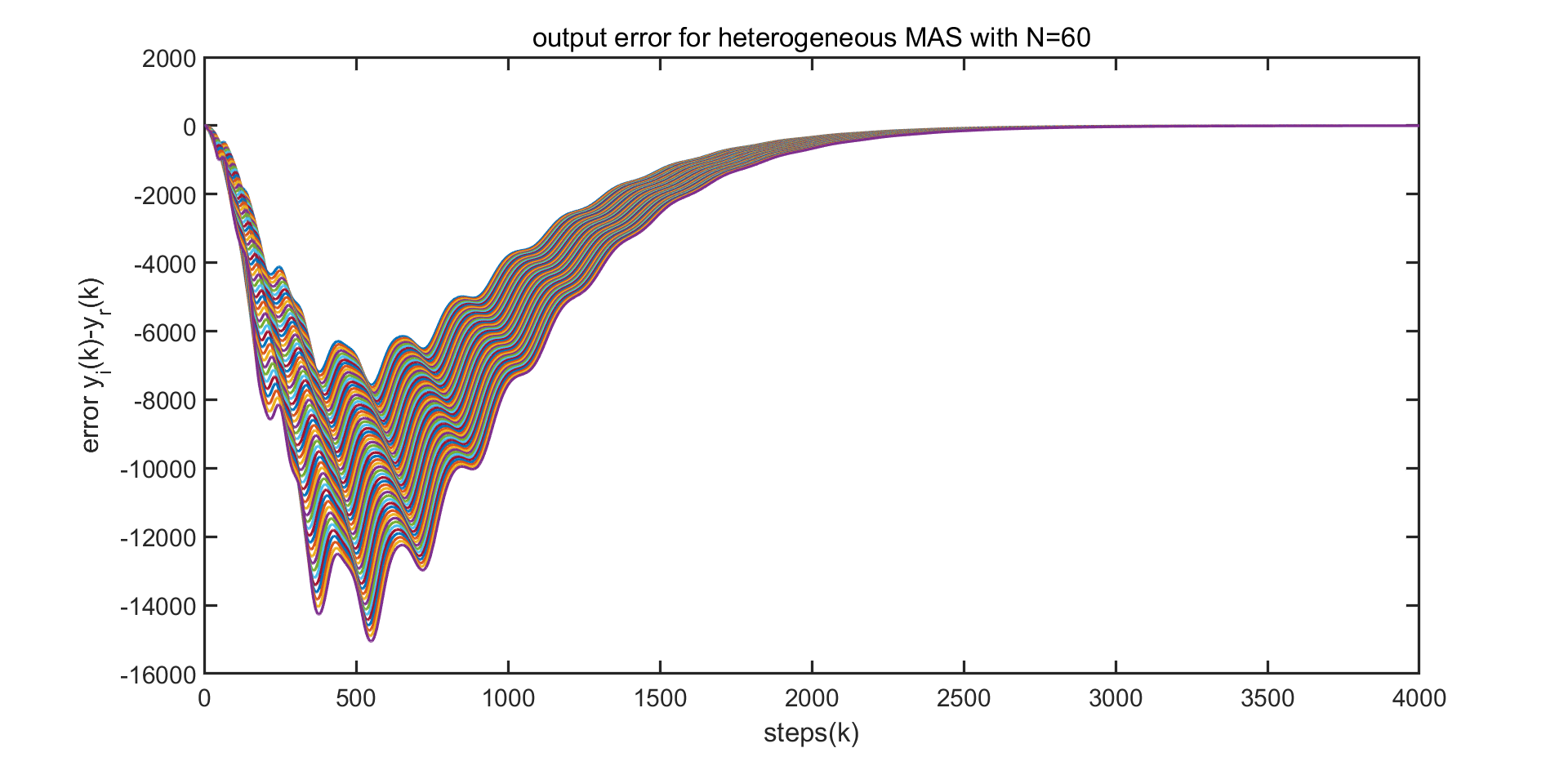}
			\centering
			\caption{Scalable exact regulated output synchronization for heterogeneous discrete-time
				MAS with communication topology described in Case II}\label{het_N60}
		\end{figure}
		
		Figures \ref{het_N6} and \ref{het_N60}  show that scalable regulated output synchronization is achieved for a heterogeneous MAS with both $N=6$ agents and $N=60$ agents. 
		
		\section{Conclusion}
		In this paper, the scalable exact output and regulated output synchronizations have been achieved for homogeneous MAS with non-introspective agents and heterogeneous MAS with introspective agents, respectively, both under the affect of disturbances and measurement noises, but with known frequencies. The \emph{scalable} collaborative protocol is developed merely upon agent models, without utilizing localized collaborative information exchange.

%
%
%
		
		\section*{Appendices}
		\appendix
		\section{The proof about the stabilizability of ($\tilde{A},\tilde{B}$)}\label{app01}
		Firstly, 
		\[
		\tilde{A}=\begin{pmatrix}
			A&B\Gamma_p\\0&S_p
		\end{pmatrix} \text{ and }\tilde{B}=\begin{pmatrix}
			BD_p\\B_p
		\end{pmatrix}.
		\]
		
		To prove the stabilizability of ($\tilde{A},\tilde{B}$), we use
		\begin{equation*}
			\rank\begin{pmatrix}
				\lambda I-A&-B\Gamma_p&-BD_p\\
				0&\lambda I-S_p&B_p
			\end{pmatrix}
			=\rank\begin{pmatrix}
				\lambda I-A&-B&0\\
				0&0&I_{n_\omega}
			\end{pmatrix}\begin{pmatrix}
				I_n&0&0\\
				0&\Gamma_p&D_p\\
				0&\lambda I-S_p&B_p
			\end{pmatrix}
		\end{equation*}
		
		Since Compensator \eqref{compensator1a} is invertible and minimum-phase, we can obtain that 
		\[
		\begin{pmatrix}
			\Gamma_p&D_p\\
			\lambda I-S_p&B_p
		\end{pmatrix}
		\]
		is invertible for all $\lambda$ in closed right-half plane.
		
		Hence, we have
		\begin{align*}
			&\rank\begin{pmatrix}
				\lambda I-A&-B\Gamma_p&-BD_p\\
				0&\lambda I-S_p&B_p
			\end{pmatrix}\\
			=&\rank\begin{pmatrix}
				\lambda I-A&-B&0\\
				0&0&I_{n_\omega}
			\end{pmatrix}\\
			=&\rank\begin{pmatrix}
				\lambda I-A&-B
			\end{pmatrix}+n_\omega\\
			=&n+n_\omega
		\end{align*}
		for all $\lambda$ in closed right-half plane due to $(A, B)$ stabilizable. This yields the required stabilizability.

		\section{The proof about the detectability of ($\tilde{C},\tilde{A}$)}\label{app1}
		
		Let $\tilde{A}_o=\begin{pmatrix}
			A&-E\\0&S
		\end{pmatrix}$ and $\tilde{C}_o=\begin{pmatrix}
			C&-G
		\end{pmatrix}$. According to Assumption A1.4, we have $\left(\tilde{C}_o, \tilde{A}_o
		\right)$ is detectable.

		Then, let $\Phi=\begin{pmatrix}
			I_n&\Pi\\0&I_{n_\omega}
		\end{pmatrix}$, we have
		\[
		\Phi\tilde{A}_o\Phi^{-1}=\begin{pmatrix}
			A&B\Gamma\\0&S
		\end{pmatrix}, \qquad \tilde{C}_o\Phi^{-1}=\begin{pmatrix}
			C&0
		\end{pmatrix}.
		\]
		Hence,
		\[
		\left(\begin{pmatrix}
			C&0
		\end{pmatrix}, \begin{pmatrix}
			A&B\Gamma\\0&S
		\end{pmatrix}\right)
		\]
		is detectable.

		\section{The proof about the detectability of ($\check{C}_i,\check{A}_i$) and ($\check{C}_i^m,\check{A}_i$)}\label{app11}
		
		Let
		\[
		\check{A}_i=\begin{pmatrix}
			\tilde{A}_{q,i}&-\tilde{E}_{q,i}\\0&S_i
		\end{pmatrix}, \check{C}_i=\begin{pmatrix}
			\tilde{C}_{q,i}& -G_i
		\end{pmatrix}, \text{ and }\check{C}_i^m=\begin{pmatrix}
			\tilde{C}_{q,i}^m&0
		\end{pmatrix}.
		\]
		Meanwhile, we have
		\begin{equation*}
			\check{A}_i=\begin{pmatrix}
				A_i&B_iC_{q,i}&-{E}_{i}\\0&A_{q,i}&0\\0&0&S_i
			\end{pmatrix}, \ \check{C}_i=\begin{pmatrix}
				{C}_{i}&0& -G_i
			\end{pmatrix}, \text{ and }\check{C}_i^m=\begin{pmatrix}
				\tilde{C}_{q,i}^m& 0&0
			\end{pmatrix}.
		\end{equation*}
		
		Since $A_{q,i}$ is Schur stable and $\left(\begin{pmatrix}
			{C}_{i}& -G_i
		\end{pmatrix}, \begin{pmatrix}
			A_i&-{E}_{i}\\0&S_i
		\end{pmatrix}\right)$ is detectable by Assumption \ref{ass1}, it implies that ($\check{C}_i,\check{A}_i$) is detectable.

		Similarly, we also obtain that ($\check{C}_i^m,\check{A}_i$) is detectable.

		\section{The proof about the detectability of ($\tilde{C}_i^m,\tilde{A}_i$)}\label{app2}

		From Appendix \ref{app11}, we have 
		\[
		\left(\begin{pmatrix} \tilde{C}_{q,i}^m& 0\end{pmatrix}, \begin{pmatrix}
			\tilde{A}_{q,i}&-\tilde{E}_{q,i}\\0&S_i
		\end{pmatrix}\right)
		\]
		is detectable.
		
		Similar to the proof of Appendix \ref{app1}, let $\Phi_i=\begin{pmatrix}
			I_n&\Pi_i\\0&I_{n_\omega}
		\end{pmatrix}$. Then, we can obtain
		\[
		\left(\begin{pmatrix} \tilde{C}_{q,i}^m& 0\end{pmatrix}\Phi^{-1}, \Phi\begin{pmatrix}
			\tilde{A}_{q,i}&-\tilde{E}_{q,i}\\0&S_i
		\end{pmatrix}\Phi^{-1}\right)
		\]
		is detectable, which is equal to 
		\[
		\left(\begin{pmatrix} \tilde{C}_{q,i}^m& -\tilde{C}_{q,i}^m\Pi_i\end{pmatrix}, \begin{pmatrix}
			\tilde{A}_{q,i}&\tilde{B}_{q,i}\Gamma_i\\0&S_i
		\end{pmatrix}\right)
		\]
		i.e., ($\tilde{C}_i^m,\tilde{A}_i$) is detectable.
		
		Using the same transformation as above detectability of 
		\[
		\left(\begin{pmatrix} \tilde{C}_{q,i}^m& -G\end{pmatrix}, \begin{pmatrix}
			\tilde{A}_{q,i}&-\tilde{E}_{q,i}\\0&S_i
		\end{pmatrix}\right)
		\]
		(see Appendix \ref{app11}) implies  detectability of ($\tilde{C}_i, \tilde{A}_i$).

\bibliographystyle{plain}
\bibliography{referenc}

\end{document}